\documentclass[twocolumn,showpacs,preprintnumbers,amsmath,amssymb,prb,aps,superscriptaddress]{revtex4-1}

\usepackage{graphicx}
\usepackage{multirow}
\usepackage{amsmath}
\usepackage{amssymb}
\usepackage{amsfonts}
\usepackage{soul}
\usepackage{verbatim}
\usepackage{bm}
\usepackage{outlines}
\usepackage{siunitx}
\usepackage[space]{grffile}
\usepackage[colorlinks=true, linkcolor=blue, citecolor=gray]{hyperref}
\usepackage[capitalize]{cleveref}
\usepackage{xr}
\usepackage{color}
\usepackage{tikz}
\usepackage{fancyhdr}
\usepackage{titlesec}

\usetikzlibrary{calc, positioning}
\newcommand{\GA}{\mathcal{G}_A}
\newcommand{\EA}{E_A}
\newcommand{\VA}{V_A}
\newcommand{\GS}{\mathcal{G}_S}
\newcommand{\ES}{E_S}
\newcommand{\VS}{V_S}
\renewcommand{\AA}{A_A}
\newcommand{\AS}{A_S}

\newcommand{\ppercycle}{\%_\mathrm{c}}

\newcommand{\ea}{\epsilon_{\mathrm{anc}}}

\newcommand{\ephys}{\epsilon_{\mathrm{phys}}}
\newcommand{\emwpm}{\epsilon_{\mathrm{L}}^{\mathrm{(MWPM)}}}

\newcommand{\ecp}{\Lambda_{\mathrm{t}}}
\newcommand{\powm}{\gamma_{\mathrm{m}}}
\newcommand{\powc}{\gamma_{\mathrm{c}}}

\newcommand{\eseventeen}{\epsilon_{\mathrm{L}}^{(17)}}
\newcommand{\efortynine}{\epsilon_{\mathrm{L}}^{(49)}}

\newcommand{\Tcorr}{\tau_{\mathrm{c}}}

\newcommand{\Nmat}{N_{\mathrm{mat}}}

\newcommand{\etad}{\eta_{\mathrm{d}}}

\newcommand{\tcycle}{\tau_{\mathrm{cycle}}}

\newcommand{\fid}{\mathcal{F}_{\mathrm{L}}}
\newcommand{\fidphys}{\mathcal{F}_{\mathrm{phys}}}

\newcommand{\eL}{\epsilon_{\mathrm{L}}}

\newcommand{\paxis}{p_{\mathrm{axis}}}
\newcommand{\pplane}{p_{\mathrm{plane}}}
\newcommand{\ero}{\epsilon_{\mathrm{RO}}}
\newcommand{\kphoton}{\kappa}
\newcommand{\disp}{\chi/\pi}

\newcommand{\Tgone}{\tau_{\mathrm{g,1Q}}}
\newcommand{\Tgtwo}{\tau_{\mathrm{g,2Q}}}
\newcommand{\Tmeas}{\tau_{\mathrm{m}}}
\newcommand{\Tdep}{\tau_{\mathrm{d}}}

\newcommand{\butterfly}[3]{\epsilon_{#1}^{#2, #3}}
\newcommand{\meas}[1]{m_{\mathrm{#1}}}

\newcommand{\fq}{f_{\mathrm{q}}}
\newcommand{\fqmax}{f_{\mathrm{q, max}}}

\newcommand{\rms}{\mathrm{rms}}
\newcommand{\twoq}{2\mathrm{Q}}
\newcommand{\oneq}{1\mathrm{Q}}
\newcommand{\Phic}{\Phi_{\mathrm{c}}}

\newcommand{\pphoton}{p_{\phi,\mathrm{photon}}}

\newcommand{\ket}[1]{\left\lvert #1 \right\rangle}
\newcommand{\Tone}{T_{1}}
\newcommand{\Ttwo}{T_{2}}
\newcommand{\Tphi}{T_{\phi}}
\newcommand{\us}{\mu\mathrm{s}}
\newcommand{\ns}{\mathrm{ns}}
\newcommand{\ms}{\mathrm{ms}}
\newcommand{\MHz}{\mathrm{MHz}}
\newcommand{\EC}{E_{\mathrm{C}}}
\newcommand{\Phio}{\Phi_0}

\newcommand{\Hz}{\mathrm{Hz}}
\newcommand{\rad}{\mathrm{rad}}

\makeatletter
\newcommand\footnoteref[1]{\protected@xdef\@thefnmark{\ref{#1}}\@footnotemark}
\makeatother

\makeatletter
\gdef\@ptsize{0}% 10pt documents
\let\@currsize\normalsize
\makeatother
\usepackage{setspace}
\begin{document}

\title{Density-matrix simulation of small surface codes under current and projected experimental noise}
\author{T.~E.~O'Brien$^*$}
\affiliation{Instituut-Lorentz, Universiteit Leiden, P.O. Box 9506, 2300 RA Leiden, The Netherlands}
\thanks{These authors contributed equally to this work.}
\author{B.~Tarasinski$^*$}
\affiliation{Instituut-Lorentz, Universiteit Leiden, P.O. Box 9506, 2300 RA Leiden, The Netherlands}
\affiliation{QuTech, Delft University of Technology, P.O. Box 5046, 2600 GA Delft, The Netherlands}
\author{L.~DiCarlo}
\affiliation{QuTech, Delft University of Technology, P.O. Box 5046, 2600 GA Delft, The Netherlands}
\affiliation{Kavli Institute of Nanoscience, Delft University of Technology, P.O. Box 5046, 2600 GA Delft, The Netherlands}
\date{\today}
\begin{abstract}
We present a density-matrix simulation of the quantum memory and computing performance of the distance-3 logical qubit Surface-17, following a recently proposed quantum circuit and using  experimental error parameters for transmon qubits in a planar circuit QED architecture. We use this simulation to optimize components of the QEC scheme (e.g., trading off stabilizer measurement infidelity for reduced cycle time) and to investigate the benefits of feedback harnessing the fundamental asymmetry of relaxation-dominated error in the constituent transmons.
A lower-order approximate calculation extends these predictions to the distance-$5$ Surface-49.
These results clearly indicate error rates below the fault-tolerance threshold of surface code, and the potential for Surface-17 to perform beyond the break-even point of quantum memory.
However, Surface-49 is required to surpass the break-even point of computation at state-of-the-art qubit relaxation times and readout speeds.
\end{abstract}
\maketitle

\section{Introduction}
Recent experimental demonstrations of small quantum simulations~\cite{OMalley16,Barends15,Langford16} and quantum error correction (QEC)~\cite{Kelly15,Riste15,Corcoles15,Ofek16} position superconducting circuits for targeting quantum supremacy~\cite{Boixo16} and quantum fault tolerance~\cite{Martinis15}, two outstanding challenges for all quantum information processing platforms. On the theoretical side, much modeling of QEC codes has been made to determine fault-tolerance threshold rates in various models~\cite{Fowler12,Landahl11,Yoder16} with different error decoders~\cite{Tomita14,Fowler09,Heim16}. However, the need for computational efficiency has constrained many previous studies to oversimplified noise models, such as depolarizing and bit-flip noise channels. This discrepancy between theoretical descriptions and experimental reality compromises the ability to predict the performance of near-term QEC implementations, and offers limited guidance to the experimentalist through the maze of parameter choices and trade-offs. In the planar circuit quantum electrodynamics (cQED)~\cite{Blais04} architecture, the major contributions to error are transmon qubit relaxation, dephasing from flux noise and resonator photons leftover from measurement, and leakage from the computational space, none of which are well-approximated by depolarizing or bit-flip channels. Simulations with more complex error models are now essential to accurately pinpoint the leading contributions to the logical error rate in the small-distance surface codes~\cite{Fowler12,Tomita14,Horsman12} currently pursued by several groups worldwide.

In this paper, we perform a density-matrix simulation of the distance-3 surface code named Surface-$17$, using the concrete quantum circuit recently proposed in~\cite{Versluis16} and the measured performance of current experimental multi-transmon cQED platforms~\cite{Bultink16,Rol16,Asaad16,Walter17}.
For this purpose, we have developed an open-source density-matrix simulation package named quantumsim~\footnote{Please visit https://github.com/brianzi/quantumsim}. We use quantumsim to extract the logical error rate per QEC cycle, $\eL$. This metric allows us to optimize and trade off between QEC cycle parameters, assess the merits of feedback control,  predict gains from future improvements in physical qubit performance, and quantify decoder performance. We compare an algorithmic decoder using minimum-weight perfect matching (MWPM) with homemade weight calculation to a simple look-up table (LT) decoder, and weigh both against an
upper bound (UB) for decoder performance obtainable from the density-matrix simulation. Finally, we make a low-order approximation to extend our predictions to the distance-$5$ Surface-$49$. The combination of results for Surface-17 and -49 allows us to make statements about code scaling and to predict the code size and physical qubit performance required to achieve break-even points for memory and computational performance.

\section{Results}
\subsection{Error rates for Surface-17 under current experimental conditions}

To quantify the performance of the logical qubit, we first define a test experiment to simulate.
Inspired by the recent experimental demonstration of distance-3 and -5 repetition codes~\cite{Kelly15}, we first focus on the performance of the logical qubit as a quantum memory.
Specifically, we quantify the ability to hold a logical $\ket{0}$ state, by initializing this state, holding it for $k\in\left\{1,\ldots,20\right\}$ cycles, performing error correction, and determining a final logical state (see Fig.~\ref{Fig_schematic} for details).
The logical fidelity $\fid[k]$ is then given by the probability to match the initial state.
We observe identical results when using $\ket{1}$ or $\ket{\pm}=\frac{1}{\sqrt{2}}(\ket{0}\pm\ket{1})$ in place of $\ket{0}$.

\begin{figure}
    \includegraphics[width=\columnwidth]{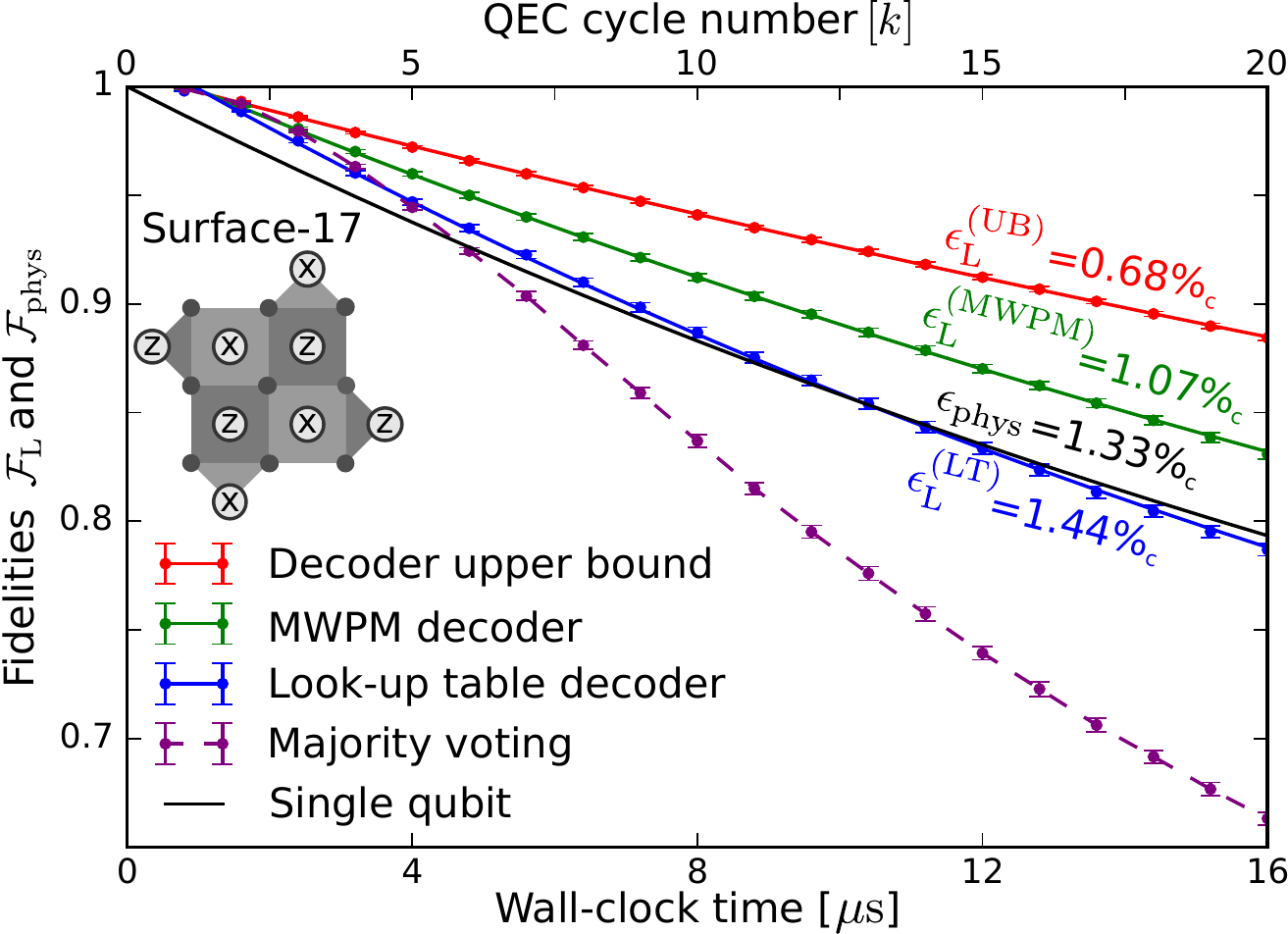}
    \caption{\label{fig:logical_fidelity}Logical fidelity $\fid[k]$ of Surface-$17$ with
        current experimental parameters (Table~\ref{table:parameters} and~\cite{Suppmaterial}),
        simulated with quantumsim as described in Fig.~\ref{Fig_schematic}.
        The results from a MWPM decoder (green) and an implementation of the LT
        decoder of~\cite{Tomita14} (blue) are compared to the decoder upper
        bound (red).  The labeled error rate is obtained from the best fit to Eq.~\eqref{eq:good_decay} (also plotted).  A further comparison is
        given to majority voting (purple, dashed), which ignores the
        outcome of individual stabilizer measurements, and to the fidelity $\fidphys$ of a single transmon
        (black) [Eq.~\eqref{eq:physical_fidelity_decay}].  Error bars ($2$ s.d.) are
        obtained by bootstrapping.} 
\end{figure}

We base our error model for the physical qubits on current typical experimental performance for transmons in planar cQED, using parameters from the literature and in-house results (e.g., gate-set tomography  measurements).
These are summarized in Table~\ref{table:parameters}, and further detailed in~\cite{Suppmaterial}.
We focus on the QEC cycle proposed in~\cite{Versluis16}, which pipelines the execution of $X$- and $Z$-type stabilizer measurements.
Each stabilizer measurement consists of three parts: a coherent step (duration $\Tcorr=2\Tgone +4\Tgtwo$), measurement ($\Tmeas$), and photon depletion from readout resonators ($\Tdep$), making the QEC cycle time $\tcycle = \Tcorr + \Tmeas + \Tdep$.

Simulating this concrete quantum circuit with the listed parameters using quantumsim, we predict $\fid[k]$ of Surface-$17$ (Fig.~\ref{fig:logical_fidelity}).
We show $\fid[k]$ for both a homemade MWPM decoder (green, described in~\cite{Suppmaterial}), and an implementation of the LT decoder of~\cite{Tomita14} (blue, described in~\cite{Suppmaterial}).
To isolate decoder performance, we can compare the achieved fidelity to an upper bound extractable from the density-matrix simulation (red, described in Sec.~\ref{sec:quantumsim}).
To assess the benefit of QEC, we also compare to a single decohering transmon, whose fidelity is calculated by averaging over the six cardinal points of the Bloch sphere:
\begin{equation}
    \label{eq:physical_fidelity_decay}
    \fidphys(t) = \tfrac{1}{6} \left( 1 + e^{-t/\Tone} \right) + \tfrac{1}{3} \left(1 + e^{-t(1/2\Tone + 1/\Tphi)} \right).
\end{equation}
The observation of $\fid[k] > \fidphys(k\tcycle)$ for large
$k$ would constitute a demonstration of QEC beyond the quantum memory break-even
point~\cite{Ofek16}. Equivalently, one can extract a logical error rate $\eL$ from a best fit
to $\fid[k]$ (as derived in Sec.~\ref{sec:protocol} as the probability of an odd number of errors occurring),
\begin{equation}
\fid[k]=\frac{1}{2}[1+(1-2\eL)^{k-k_0}].\label{eq:good_decay}
\end{equation}
Here, $k_0$ and $\eL$ are the parameters to be fit. We compare $\eL$ to the physical error rate
\begin{equation}
\ephys = -\tcycle \left. \frac{d \fidphys(t)}{dt}\right|_{t=0}=\frac{\tcycle}{3\Tone}+\frac{\tcycle}{3\Tphi}.
\end{equation}
We observe $\eL=1.44\,\ppercycle$ for the LT decoder, $\eL=1.07\,\ppercycle$ for the MWPM decoder, and $\eL=0.68\,\ppercycle$ at the decoder upper bound ($\ppercycle$ = $\%$ per cycle).
The latter two fall below $\ephys=1.33\,\ppercycle$.
Defining the decoder efficiency $\etad = \eL^{\mathrm{(UB)}}/\eL$, we find
$\etad^{\mathrm{(LT)}} = 0.47$ and $\etad^{\mathrm{(MWPM)}} = 0.64$.

We can also compare the multi-cycle error correction to majority voting, in which the state declaration is based solely on the output of the final data
qubit measurements (ancilla measurements are ignored).
Majority voting corrects any single data qubit error (over the entire experiment), and thus exhibits a
quadratic decay for small $k$~\footnote{A distance-$d$ code with majority voting alone should exhibit a $(d+1)/2$-order decay}.
A decoder should also be able to correct (at least) a single error, and thus should produce the same behavior at low $k$, delaying the onset of exponential decay in $\fid[k]$.
In fact, a good test for the performance of a MWPM decoder is to ensure it can outperform the majority vote at short timescales, as suboptimal configuration will prevent this (as seen for the look-up table decoder).

With the baseline for current performance established, we next investigate $\eL$ improvements that may be achieved by two means.
First, we consider modifications to the QEC cycle  at fixed physical performance.
Afterwards, we consider the effect of improving physical qubit $\Tone$ and $\Tphi$.

\subsection{Optimization of logical error rates with current experimental conditions}
\begin{figure}
\includegraphics[width=\columnwidth]{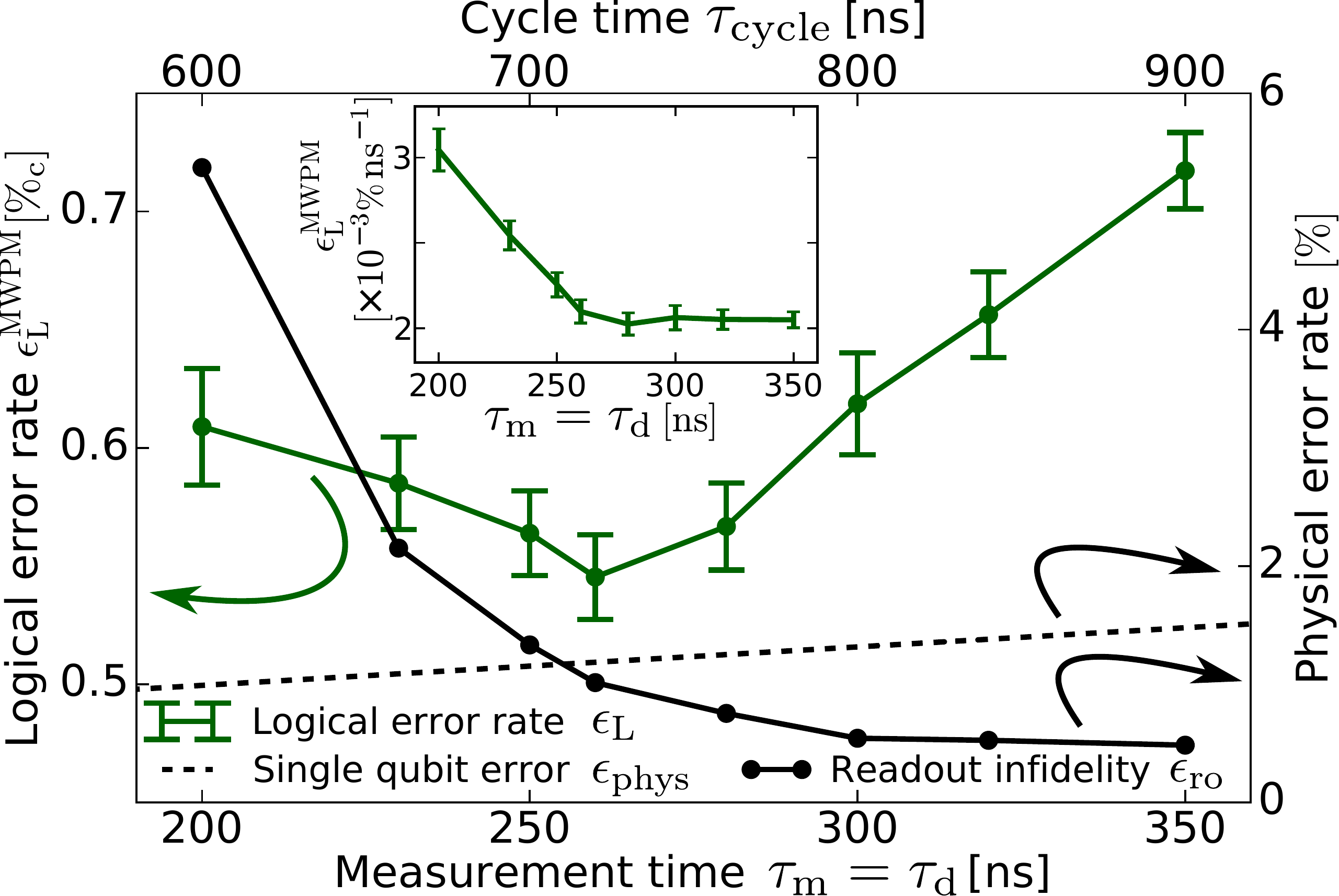}
\caption{\label{Fig_2_Measurement_time}Optimization of the logical error rate (per cycle) of Surface-$17$ as a function of measurement-and-depletion time~\cite{Bultink16}.
    Changes in the underlying physical error rates are shown as well.
    Decreasing the measurement time causes an increase in the readout infidelity (solid black curve with dots), whilst decreasing the single qubit decay from $\Tone$ and $\Ttwo$ (black dashed curve) for all qubits.
    The logical rate with an MWPM decoder (green curve) is minimized when these error rates are appropriately balanced.
    The logical error rate is calculated from the best fit of Eq.~\eqref{eq:good_decay}. Error bars ($2$ s.d.) are obtained by bootstrapping ($N=10,000$ runs).
    Inset: Logical error rate per unit time, instead of per cycle.
}
\end{figure}

Error sources in current cQED setups derive primarily from transmon decoherence, as opposed to gate and measurement errors produced by control electronics.
Thus, a path to reducing $\eL$ may be to decrease $\tcycle$.
Currently, the cycle is dominated by $\Tmeas + \Tdep$.
At fixed readout power, reducing $\Tmeas$ and $\Tdep$ will reduce $\tcycle$
at the cost of increased readout infidelity $\ero$ (described in Sec.~\ref{sec:measurement}).
We explore this trade-off in Fig.~\ref{Fig_2_Measurement_time}, using a
linear-dispersive readout model \cite{FriskKockum12}, keeping $\Tmeas = \Tdep$
and assuming no leftover photons.
Because of the latter, $\eL^{\mathrm{(MWPM)}}$ reduces from $1.07\,\ppercycle$~(Fig.~\ref{fig:logical_fidelity}) to $0.62\,\ppercycle$ at
$\Tmeas = 300\,\text{ns}$.
The minimum $\emwpm = 0.55\,\ppercycle $ is achieved at around $\Tmeas=260~\ns$. This is perhaps counterintuitive, as
$\ephys$ reduces only $0.13\,\ppercycle$ while $\ero$ increases $0.5\,\%$. However, it reflects the different sensitivity of the code to different types of errors.
Indeed, $\emwpm$ is smaller for $\Tmeas=200~\ns$ than for $\Tmeas=300~\ns$, even though $\ero$ increases to $5\,\%$. It is interesting to note that the optimal $\Tmeas$ for quantum memory, which
minimizes logical error per unit time, rather than per cycle, is $\Tmeas =
280\,\text{ns}$ (Fig.~\ref{Fig_2_Measurement_time} inset). This shows that different cycle parameters might be optimal for computation and memory applications.

Next, we consider the possibility to reduce $\eL$ using feedback control.
Since $\Tone$ only affects qubits in the excited state, the error rate of ancillas in Surface-17 is roughly two times higher when in the excited state.
The unmodified syndrome extraction circuit flips the ancilla if the corresponding stabilizer value is -1, and since ancillas are not reset between cycles, they will spend significant amounts of time in the excited state.
Thus, we consider using feedback to hold each ancilla in the ground state as much as possible.
We do not consider feedback on data qubits, as the highly entangled logical states are equally susceptible to $\Tone$. 

\begin{figure}
\includegraphics[width=\columnwidth]{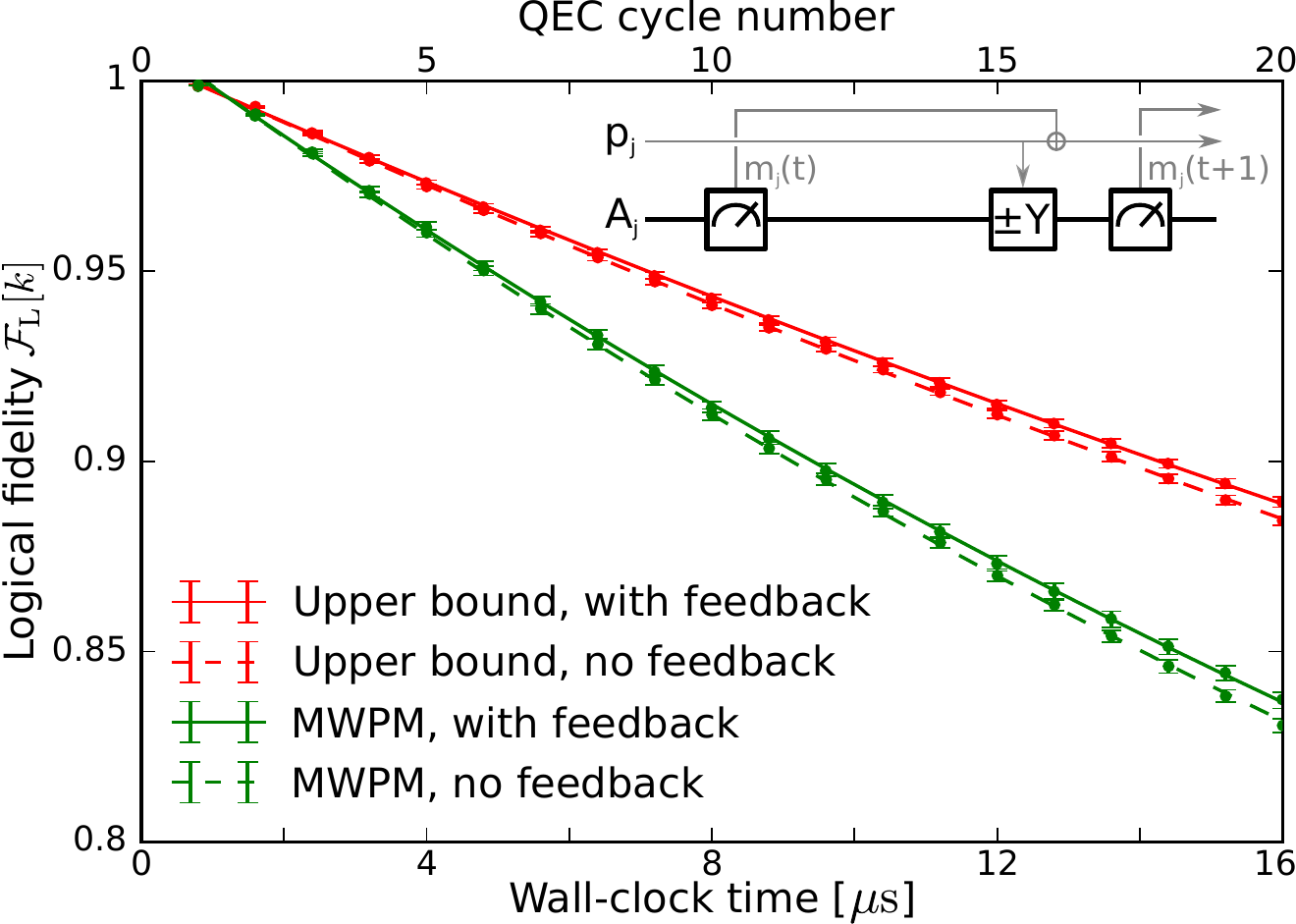}
\caption{\label{Fig_4_FeedbackFidelity}Logical fidelity of Surface-$17$ with (solid) and without (dashed) an additional feedback scheme.
    The performance of a MWPM decoder (green) is compared to the decoder upper bound (red).
    Curves are fits of Eq.~\eqref{eq:good_decay} to the data, and error bars ($2$ s.d.) are given by bootstrapping, with each point averaged over $10,000$ runs.
    Inset: Method for implementing the feedback scheme.
    For each ancilla qubit $A_j$, we store a parity bit $p_j$, which decides the sign of the $R_y(\pi/2)$ rotation at the end of each coherent step.
    The time $A_j$ spends in the ground state is maximized when $p_j$ is updated each cycle $t$ by XORing with the measurement result from cycle $t-1$, after the rotation of cycle $t$ has been performed.
}
\end{figure}

The feedback scheme (Inset of Fig. 3) consists of replacing the $R_y(\pi/2)$ gate at the end of the coherent step with a $R_y(-\pi/2)$ gate for some of the ancillas, depending on a classical control bit $p$ for each ancilla.
  This bit $p$ represents an estimate of the stabilizer value, and the ancilla is held in the ground state whenever this estimate is correct (i.e.~in the absence of errors).
Figure~\ref{Fig_4_FeedbackFidelity} shows the effect of this feedback on the logical fidelity, both for the MWPM decoder and the decoder upper bound.
We observe $\eL$ improve only $0.05\,\ppercycle$ in both cases.
Future experiments might opt not to pursue these small gains in view of the technical challenges added by feedback control.

\subsection{Projected improvement with advances in quantum hardware}\label{sec:analytics}
\begin{figure}
  \includegraphics[width=\columnwidth]{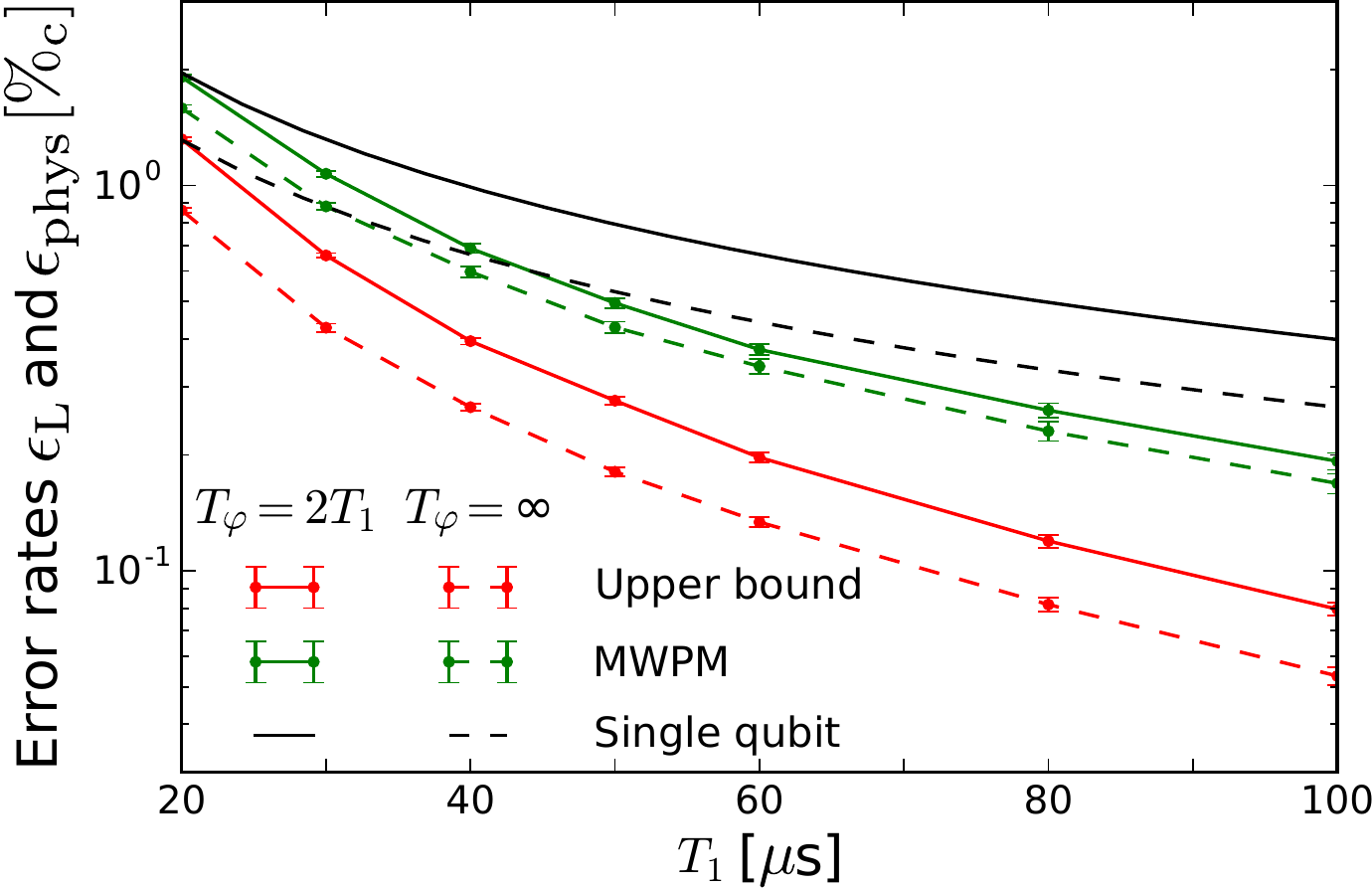}
  \caption{\label{Fig_3_T1}$\Tone$ dependence of the Surface-$17$ logical error rate (MWPM and UB) and the physical error rate.
  We either fix $\Tphi=2\Tone$ (solid) or $\Tphi=\infty$ (dashed).
  Logical error rates are extracted from a best fit of Eq.~\eqref{eq:good_decay} to $\fid[k]$ over $k=1,\ldots,20$ QEC cycles, averaged over $N=50,000$ runs. Error bars ($2$ s.d.) are calculated by bootstrapping.
}
\end{figure}

We now estimate the performance increase that may result from improving the transmon relaxation and dephasing times via materials and filtering improvements.
To model this, we return to $\tcycle=800~\ns$, and adjust $\Tone$ values with both $\Tphi=2\Tone$ (common in experiment) and $\Tphi=\infty$ (all white-noise dephasing eliminated).
We retain the same rates for coherent errors, readout infidelity, and photon-induced dephasing as in Fig.~\ref{fig:logical_fidelity}.
Figure~\ref{Fig_3_T1} shows the extracted $\eL$ and $\ephys$ over the $\Tone$ range covered.
For the MWPM decoder (upper bound) and $\Tphi=2\Tone$, the memory figure of merit $\powm=\ephys/\eL$ increases from $1.3$ $ (2)$ at $\Tone = 30~\us$ to $2$ $(5)$ at $100~\us$.
Completely eliminating white-noise dephasing will increase $\powm$ by $10\%$ with MWPM and $30\%$ at the upper bound.

\begin{figure}
\includegraphics[width=\columnwidth]{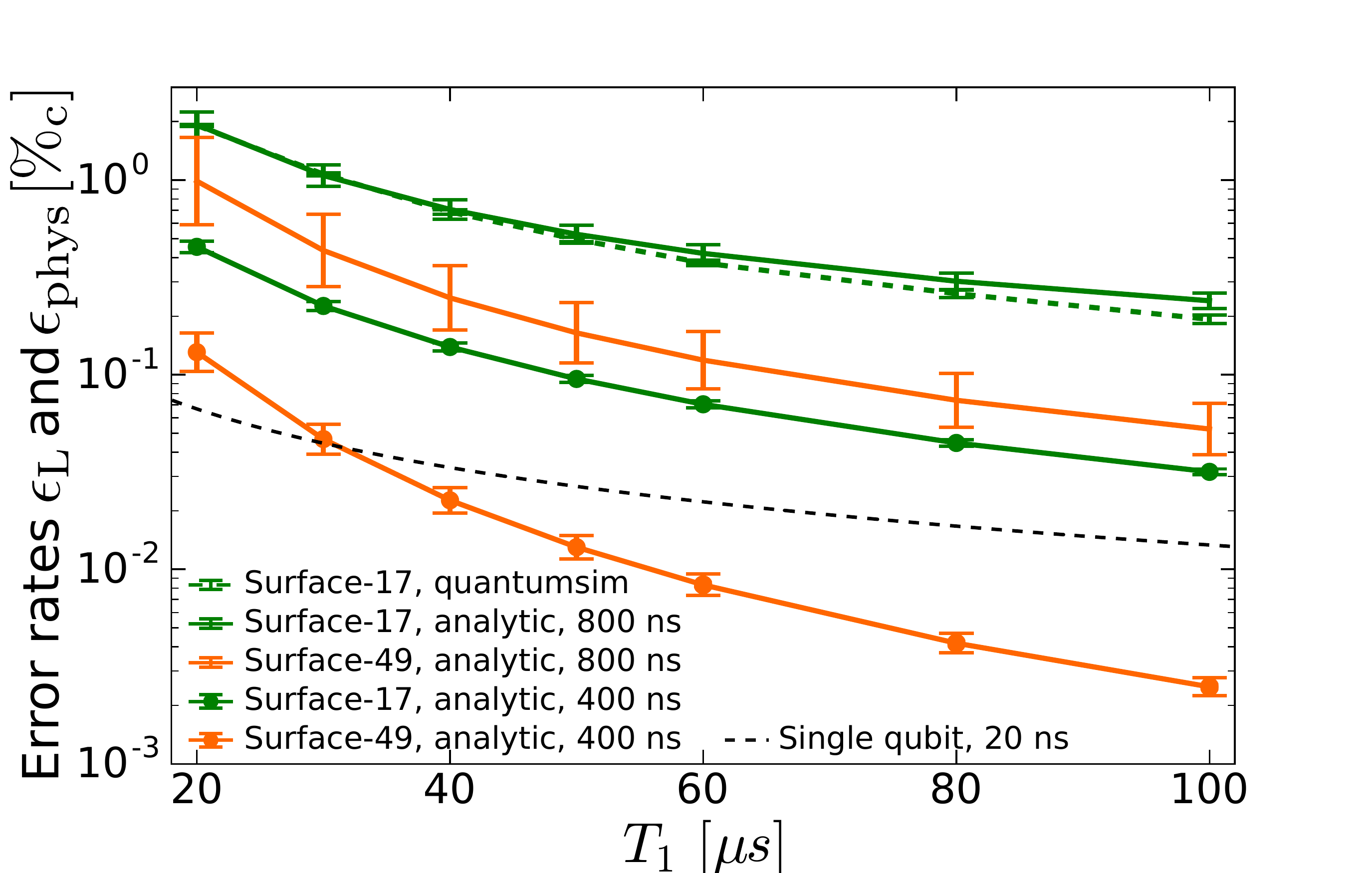}
\caption{\label{Fig_6_Analytic_extension_49}Analytic approximation of $\eL$ for Surface-$17$ (green) and Surface-$49$ (orange) using a MWPM decoder.
    Details of the calculation of points and error bars are given in~\cite{Suppmaterial}.
    All plots assume $\Tphi=2\Tone$, and $\tcycle=800~\ns$ (crosses) or $400~\ns$ (dots).
    Numerical results for Surface-$17$ with $\tcycle=800~\ns$ are also plotted for comparison (green, dashed).
    The physical-qubit computation metric is given as the error incurred by a single qubit over the resting time of a single-qubit gate (black, dashed).
}
\end{figure}

A key question for any QEC code is how $\eL$ scales with code distance $d$.
Computing power limitations preclude similar density-matrix simulations of the $d=5$ surface code Surface-$49$.
However, we can approximate the error rate by summing up all lowest-order error chains (as calculated for the MWPM decoder), and deciding individually whether or not these would be corrected by a MWPM decoder (see~\cite{Suppmaterial} for details).
Figure~\ref{Fig_6_Analytic_extension_49} shows the lowest-order approximation to the logical error rates of Surface-$17$ and -$49$ over a range of $\Tone=\Tphi/2$.
Comparing the Surface-$17$ lowest-order approximation to the quantumsim result shows good agreement and validates the approximation.
We observe a lower $\eL$ for Surface-49 than for -17, indicating quantum fault tolerance over the $\Tone$ range covered. The fault-tolerance figure of merit defined in~\cite{Martinis15}, $\ecp=\eseventeen/\efortynine$, increases from $2$ to $4$ as $\Tone$ grows from $30$ to $100~\us$.

As a rough metric of computational performance, we offer to compare $\eL$ (per cycle) to the error accrued by a physical qubit idling over $\Tgone$. We define a metric for computation performance,
$\powc=(\ephys\Tgone)/(\eL\tcycle)$ and $\powc=1$ as a computational break-even point. Clearly, using the QEC cycle parameters of Table~\ref{table:parameters} and even with $\Tone$ improvements, neither Surface-17 nor -49 can break-even computationally.
However, including the readout acceleration recently demonstrated in~\cite{Walter17}, which allows $\Tmeas=\Tdep=100~\ns$ and $\tcycle=400~\ns$, Surface-49 can cross $\powc=1$  by $\Tone=40~\us$. In view of first reports of $\Tone$ up to $80~\us$ emerging for planar transmons~\cite{Paik16,Gustavsson16}, this important milestone may be within grasp.

\section{Discussion}
\subsection{Computational figure of merit}

We note that our metric of computational power is not rigorous, due to the different gate sets
available to physical and logical qubits.  Logical qubits can execute multiple
logical $X$ and $Z$ gates within one QEC cycle, but require a few cycles for
two-qubit and Hadamard gates (using the proposals
of~\cite{Horsman12,Yoder16}), and state distillation over many cycles to
perform non-Clifford gates.  As such, this metric is merely a rough benchmark
for computational competitiveness of the QEC code.  However, given the amount
by which all distance-$3$ logical fidelities fall above this metric, we find it
unlikely that these codes will outperform a physical qubit by any fair
comparison in the near future.

\subsection{Decoder performance}\label{sec:decoder}
A practical question facing quantum error correction is how best to balance the trade-off between decoder complexity and performance.
Past proposals for surface-code computation via lattice surgery~\cite{Horsman12} require the decoder to provide an up-to-date estimate of the Pauli error on physical qubits during each logical $T$ gate.
Because tracking Pauli errors through a non-Clifford gate is inefficient, however implemented, equivalent requirements will hold for any QEC code~\cite{Terhal15}.
A decoder is thus required to process ancilla measurements from one cycle within the next (on average).
This presents a considerable challenge for transmon-cQED implementations, as $\tcycle < 1\,\us$.
This short time makes the use of computationally intensive decoding schemes difficult, even if they provide lower $\eL$.

The leading strategy for decoding the surface code is MWPM using the blossom algorithm of Edmonds~\cite{Fowler12,Fowler09,Fowler14}.
Although this algorithm is challenging to implement, it scales linearly in code distance~\cite{Fowler14}.
The algorithm requires a set of weights (representing the probability that two given error signals are connected by a chain of errors) as input.
An important practical question (see~\cite{Suppmaterial}) is whether these weights can be calculated on the fly, or must be precalculated and stored.
On-the-fly weight calculation is more flexible. For example, it can take into account the difference in error rates between an ancilla measured in the ground and in the excited state.
The main weakness of MWPM is the inability to explicitly detect $Y$ errors. In fact,~\cite{Suppmaterial} shows that MWPM is nearly perfect in the absence of $Y$ errors.
The decoder efficiency $\etad$ may significantly increase by extending MWPM to account for correlations between detected $X$ and $Z$ errors originating from $Y$ errors~\cite{Delfosse14,Fowler13b}.

If computational limitations preclude a  MWPM decoder from keeping up with $\tcycle$, the look-up table decoder may provide a straightforward solution for Surface-17.
However, at current physical performance, the $\etad$ reduction will make Surface-17 barely miss memory break-even (Fig.~\ref{fig:logical_fidelity}). Furthermore, memory requirements make look-up table decoding already impractical for Surface-49.
Evidently, real-time algorithmic decoding by MWPM or improved variants is an important research direction already at low code distance.

\subsection{Other observations}\label{sec:gen_obs}
The simulation results allow some further observations. Although we have focused on superconducting qubits, we surmise that the following statements are fairly general.

We observe that small quasi-static qubit errors are suppressed by the repeated measurement.
In our simulations, the $1/f$ flux noise producing $0.01$ radians of phase error per flux pulse on a qubit has a diamond norm approximately equal to the $\Tone$ noise, but a trace distance $100$ times smaller.
As the flux noise increases $\eL$ by only $0.01\,\ppercycle$, it appears $\eL$ is dependent on the trace distance rather than the diamond norm of the underlying noise components.
Quasi-static qubit errors can then be easily suppressed, but will also easily poison an experiment if unchecked.

We further observe that above a certain value, ancilla and measurement errors have a diminished effect on $\eL$.
In our error model, the leading sources of error for a distance $d$ code are chains of $(d-1)/2$ data qubit errors plus either a single ancilla qubit error or readout error, which together present the same syndrome as a chain of $(d+1)/2$ data qubit errors.
An optimal decoder decides which of these chains is more likely, at which point the less-likely chain will be wrongly corrected, completing a logical error.
This implies that if readout infidelity ($\ero$) or the ancilla error rate ($\ea$) is below the data qubit ($\ephys$) error rate, $\eL\propto(\ea+\ero)\ephys^{(d-1)/2}$.
However, if $\ero$ ($\ea$) $>\ephys$, $\eL$ becomes independent of $\ero$ ($\ea$), to lowest order.
This can be seen in Fig.~\ref{Fig_2_Measurement_time}, where the error rate is almost constant as $\ero$ exponentially increases.
This approximation breaks down with large enough $\ea$ and $\ero$, but presents a counterintuitive point for experimental design; $\eL$ becomes less sensitive to measurement and ancilla errors as these error get worse.

A final, interesting point for future surface-code computation is shown in Fig.~\ref{Fig_2_Measurement_time}: the optimal cycle parameters for logical error rates per cycle and per unit time are not the same.
This implies that logical qubits functioning as a quantum memory should be treated differently to those being used for computation.
This idea can be extended further: at any point in time, a large quantum computer performing a computation will have a set $S_m$ of memory qubits which are storing part of a large entangled state, whilst a set $S_c$ of computation qubits containing the rest of the state undergo operations.
To minimize the probability of a logical error occurring on qubits within both $S_c$ and $S_m$, the cycle time of the qubits in $S_c$ can be reduced to minimize the rest time of qubits in $S_m$.
As a simple example, consider a single computational qubit $q_c$ and a single memory qubit $q_m$ sharing entanglement.
Operating all qubits at $\tcycle=720~\ns$ to minimize $\eL$ would lead to a $1.09\%$ error rate for the two qubits combined.
However, shortening the $\tcycle$ of $q_c$ reduces the time over which $q_m$ decays.
If $q_c$ operates at $\tcycle=600~\ns$, the average error per computational cycle drops to $1.06\%$, as $q_m$ completes only $5$ cycles for every $6$ on $q_c$.
Although this is only a meager improvement, one can imagine that when many more qubits are resting than performing computation, the relative gain will be quite significant.

\subsection{Effects not taken into account}
Although we have attempted to be thorough in the detailing of the circuit, we have neglected certain effects.
We have used a simple model for C-Z gate errors as we lack data from experimental tomography (e.g.~one obtained from two-qubit gate-set tomography~\cite{Blume13}).
Most importantly, we have neglected leakage, where a transmon is excited out of the two lowest energy states, i.e., out of the computational subspace.
Previous experiments have reduced the leakage probability per C-Z gate to $\sim0.3\%$~\cite{Barends14}, and per single-qubit gate to $\sim0.001\%$~\cite{Chen16}.
Schemes have also been developed to reduce the accumulation of leakage~\cite{Fowler13}. Extending quantumsim to include and investigate leakage is a next target. However, the representation of the additional quantum state can increase the simulation effort significantly [by a factor of $(9/4)^{10} \approx 3000$].
To still achieve this goal, some further approximations or modifications to the simulation will be necessary.
Future simulations will also investigate the effect of spread in qubit parameters, both in space (i.e., variation of physical error rates between qubits) and time (e.g., $\Tone$ fluctuations), and
cross-talk effects such as residual couplings between nearest and next-nearest neighbor transmons, qubit cross-driving, and qubit dephasing by measurement pulses targeting other qubits.

\section{Methods}
\subsection{Simulated experimental procedure}
\subsubsection{Surface-$17$ basics}
A QEC code can be defined by listing the data qubits and the stabilizer measurements that are repeatedly performed upon them~\cite{GottesmanPhD}.
In this way, Surface-$17$ is defined by a $3\times 3$ grid of data qubits $\{D_0,\ldots D_8\}$.
In order to stabilize a single logical qubit, $9-1=8$ commuting measurements are performed.
The stabilizers are the weight-two and weight-four $X$- and $Z$-type parity operators $X_2 X_1$,  $Z_3 Z_0$, $X_4 X_3 X_1 X_0$, $Z_5 Z_4 Z_2 Z_1$, $Z_7 Z_6 Z_4 Z_3$, $X_8 X_7 X_5 X_4$, $Z_8 Z_5$, and $X_7 X_6$, where $X_j$ ($Z_j$) denotes the $X$ ($Z$) Pauli operator acting on data qubit $D_j$.
Their measurement is realized indirectly using nearest-neighbor interactions between data and ancilla qubits arranged in a square lattices, followed by ancilla measurements [Fig.~\ref{Fig_schematic}(a)].
This leads to a total of $17$ physical qubits when a separate ancilla is used for each individual measurement.
We follow the circuit realization of this code described in~\cite{Versluis16}, for which we give a schematic description in Fig.~\ref{Fig_schematic}(b) (see~\cite{Suppmaterial} for a full circuit diagram).

In an experimental realization of this circuit, qubits will regularly accumulate errors.
Multiple errors that occur within a short period of time (e.g., one cycle) form error `chains' that spread across the surface.
Errors on single qubits, or correlated errors within a small subregion of Surface-$17$, fail to commute with the stabilizer measurements, creating error signals that allow diagnosis and correction of the error via a decoder.
However, errors that spread across more than half the surface in a short enough period of time are misdiagnosed, causing an error on the logical qubit when wrongly corrected~\cite{Fowler12}.
The rate at which these logical errors arise is the main focus of this paper.

\subsubsection{Protocol for measurement of logical error rates}\label{sec:protocol}
\begin{figure*}
\centering{
\includegraphics[width=\textwidth]{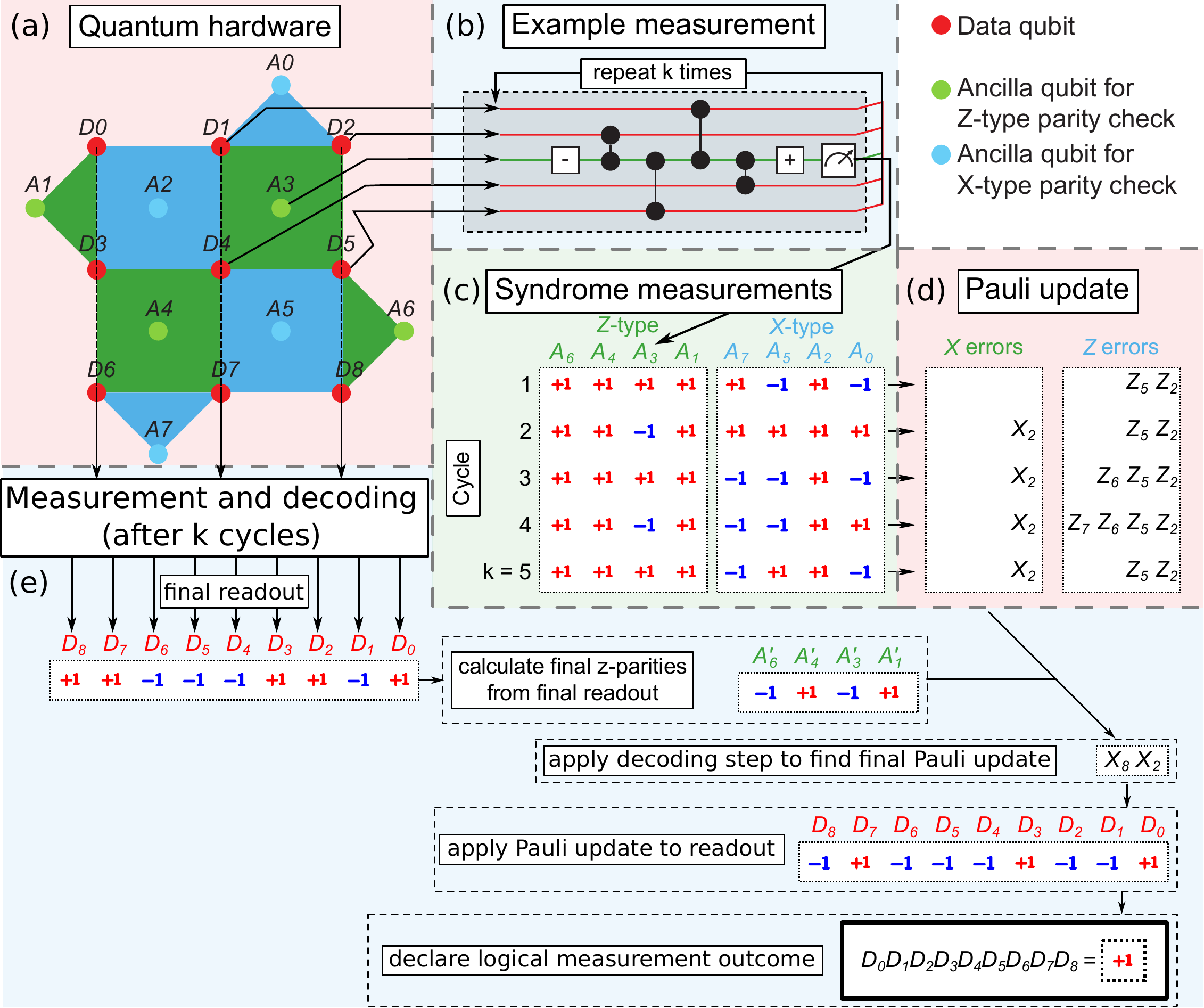}}
\caption{\label{Fig_schematic}Schematic overview of the simulated experiment.
    (a) 17 qubits are arranged in a surface code layout (legend top-right).
    The red data qubits are initialized in the ground state $\ket{0}$, and projected into an eigenstate of the measured $X$- (blue) and $Z$- (green) type stabilizer operators.
    (b) A section of the quantum circuit depicting the four-bit parity measurement implemented by the $A_3$ ancilla qubit ($+$/$-$ refer to $R_y(\pm\pi/2$) single-qubit rotations).
    The ancilla qubit (green line, middle) is entangled with the four data qubits (red lines) to measure $Z_1Z_2Z_4Z_5$.
    Ancillas are not reset between cycles. Instead, the implementation relies on the quantum non-demolition nature of measurements. The stabilizer is then the product of the ancilla measurement results of successive cycles.
    This circuit is performed for all ancillas and repeated $k$ times before a final measurement of all (data and ancilla) qubits.
    (c) All syndrome measurements of the $k$ cycles are processed by the decoder.
    (d) After each cycle, the decoder updates its internal state to represent the most likely set of errors that occurred.
    (e) After the final measurement, the decoder uses the readout from the data qubits, along with previous syndrome measurements, to declare a final logical state.  To this end,  the decoder processes the $Z$-stabilizers obtained directly from the data qubits, finalizing its prediction of most likely errors.     The logical parity is then determined as the product of all data qubit parities ($\prod_{j=0}^8D_j$) once the declared errors are corrected.
    The logical fidelity $\fid$ is the probability that this declaration is the same as the initial state ($\ket{0}$).}
\end{figure*}

As the performance measure of Surface-$17$, we study the fidelity of the logical qubit as a quantum memory.
We describe our protocol with an example `run' in Fig.~\ref{Fig_schematic}.
We initialize all qubits in $\ket{0}$ and perform $k = 1,2,\ldots,20$ QEC cycles [Fig.~\ref{Fig_schematic}(b)].
Although this initial state is not a stabilizer eigenstate, the first QEC cycle projects the system into one of the $16$ overlapping eigenstates within the $+1$ eigenspace for $Z$ stabilizers, which form the logical $\ket{0}$ state~\cite{Fowler12}.
This implies that, in the absence of errors, the first measurement of the $Z$ stabilizers will be $+1$, whilst that of the $X$ stabilizers will be random.
In the following cycles, ancilla measurements of each run [Fig.~\ref{Fig_schematic}(c)] are processed using a classical decoding algorithm.
The decoder computes a Pauli update after each QEC cycle [Fig.~\ref{Fig_schematic}(d)].
This is a best estimate of the Pauli operators that must be applied to the data qubits to transform the logical qubit back to the logical $\ket{0}$ state.
The run ends with a final measurement of all data qubits in the computational basis.
From this 9-bit outcome, a logical measurement result is declared [Fig.~\ref{Fig_schematic}(e)].
First, the four $Z$-type parities are calculated from the 9 data-qubit measurement outcomes and presented to the decoder as a final set of parity measurements.
This ensures that the final computed Pauli update will transform the measurement results into a set that measures $+1$ for all $Z$ stabilizers.
This results in one of $32$ final measurements, from which the value of a logical $Z$ operator can be calculated to give the measurement result (any choice of logical operator gives the same result).
The logical fidelity $\fid[k]$ after $k$ QEC cycles is defined as the probability of this declared result matching the initial $+1$ state.

At long times and with low error rates, Surface codes have a constant logical error rate $\eL$. The fidelity $\fid[k]$ is obtained by counting the probability of an odd number of errors having occurred in total (as two $\sigma_x$ errors cancel)~\cite{Rol16,Terhalcomment}:
\begin{align}
\fid[k]=1-\sum_{l\;\mathrm{odd}}{k \choose l}\eL^l(1-\eL)^{k-l}.
\end{align}
Here, the combinatorial factor counts the number of combinations of $l$ errors in $k$ rounds, given an $\eL$ chance of error per round. This can be simplified to
\begin{align}
\fid[k]&=1-\frac{1}{2}\sum_l{k \choose l}\eL^l(1-\eL)^{k-l}(1-(-1)^l)\nonumber\\
&=1-\frac{1}{2}\left[(1-\epsilon_L+\epsilon_L)^k-(1-\epsilon_L-\epsilon_L)^k\right] \nonumber \\
&=\frac{1}{2}[1+(1-2\eL)^{k}].\label{eq:bad_decay}
\end{align}
However, at small $k$, the decay is dominated by the majority vote, for which $\eL\propto (k \ephys)^{(d+1)/2}$.
For example, for all the Surface-$17$ decay curves, we observe a quadratic error rate at small $k$, as opposed to the linear slope predicted by Eq.~\eqref{eq:bad_decay}.
In order to correct for this, we shift the above equation in $k$ by a free parameter $k_0$, resulting in Eq.~\eqref{eq:good_decay}.
This function fits well to data with $k \ge 3$ in all plots, and thus allows accurate determination of $\eL$.

\subsubsection{The quantumsim simulation package}\label{sec:quantumsim}
Quantumsim performs calculations on density matrices utilizing a graphics processing unit in a standard desktop computer.
Ancillas are measured at the end of each cycle, and thus not entangled with the rest of the system.
As such, it is possible to obtain the effect of the QEC cycle on the system without explicitly representing the density matrix of all 17 qubits simultaneously.
The simulation is set up as follows: the density matrix of the nine data qubits is allocated in memory with all qubits initialized to $\ket{0}$.
One- and two-qubit gates are applied to the density matrix as completely positive, trace preserving maps represented by Pauli transfer matrices.
When a gate involving an ancilla qubit must be performed, the density matrix of the system is dynamically enlarged to include that one ancilla.

Qubit measurements are simulated as projective and following the Born rule, with projection probabilities given by the squared overlap of the input state with the measurement basis states.
In order to capture  empirical measurement errors, we implement a black-box measurement model (Sec.~\ref{sec:measurement}) by sandwiching the measurement between idling processes.
The measurement projects the system to a product state of the ancilla and the projected sub-block of the density matrix.
We can therefore remove the ancilla from the density matrix and only store its state right after projection, and continue the calculation with the partial density matrix of the other qubits.
Making use of the specific arrangement of the interactions between
ancillas and data qubits in Surface-17, it is possible to apply all
operations to the density matrix in such an order (shown
in~\cite{Suppmaterial}) that the total size of the density matrix never
exceeds $2^{10}\times2^{10}$ (nine data qubits plus one ancilla), which
allows relatively fast simulation. We emphasize that with the choice of error model in this work, this approach gives the same result as a full simulation on a 17-qubit density matrix. Only the introduction of residual entangling interactions between data and ancilla qubits (which we do not consider in this work) would make the latter necessary.
On our hardware (see~\cite{Suppmaterial}), simulating one QEC cycle of Surface-17 with quantumsim takes $25~\ms$.

We highlight an important advantage of doing density-matrix calculations with quantumsim.
We do not perform projective measurements of the data qubits.
Instead, after each cycle, we extract the diagonal of the data-qubit density matrix, which represents the probability distribution if a final measurement were performed.
We leave the density matrix undisturbed and continue simulation up to $k=20$.
This is a very useful property of the density-matrix approach, because having a probability distribution of all final readout events greatly reduces sampling noise.

Our measurement model includes a declaration error probability (see Sec.~\ref{sec:measurement}), where the projected state of the ancilla after measurement is
not the state reported to the decoder. Before decoding, we thus apply errors to the outcomes
of the ancilla projections, and smear the probability distribution of the data qubit measurement.
To then determine the fidelity averaged over this probability distribution, we present
all 16 possible final $Z$-type parities to the decoder.
This results in 16 different final Pauli updates, allowing us to determine correctness of the decoder for all 512 possible measurement outcomes.
These are then averaged over the simulated probability distribution.
This produces good results after about $\sim 10^4$ simulated runs.

A second highlight of quantumsim is the possibility to quantify the sub-optimality of the decoder.
The fidelity of the logical qubit obtained in these numerical simulations is a combination of the error rates of the physical qubits and the approximations made by the decoder.
Full density-matrix simulations make it possible to disentangle these two contributions.
Namely, the fidelity is obtained by assigning correctness to each of the 512 possible readouts according to 16 outputs of the decoder, and summing the corresponding probabilities accordingly.
If the probabilities are known, it is easy to determine the 16 results that a decoder should output in order to maximize fidelity (i.e., the output of the best-possible decoder). 
This allows placing a decoder upper bound $\fid^{\max}$ on logical fidelity as limited by the physical qubits independent of the decoder. Conversely, it also allows quantifying sub-optimality in the decoder used.
In fact, we can make the following reverse statement: if our measurement model did not include a declaration error, then we could use the simulation to find the final density matrix of the system conditioned on a syndrome measurement.
From this, the simulation could output exactly the 16 results that give $\fid^{\max}$, so that quantumsim could thus be used as a maximum-likelihood decoder.
In this situation, $\fid^{\max}$ would not only be an upper bound, but indeed the performance of the best-possible decoder.
However, as we add the declaration errors after simulation, we can only refer to $\fid^{\max}$ as the decoder upper bound.

\subsection{Error models}
We now describe the error model used in the simulations. 
Our motivation for the development of this error model is to provide a limited number of free parameters to study, whilst remaining as close to known experimental data as possible. 
As such, we have taken well-established theoretical models as a base, and used experimental tomography to provide fixed parameters for observed noise beyond these models.
The parameters of the error model are provided in~\cite{Suppmaterial}.

\begin{table}[h]
\begin{tabular}{| l | l | l|l| }
    \hline
    Parameter               & Symbol                    & Value                 & Reference                 \\
    \hline
    Qubit relaxation time                                   & $\Tone$                   & $30~\us$              &~\cite{Bultink16}          \\
    Qubit dephasing time (white noise)                      & $\Tphi$                   & $60~\us$              &~\cite{Asaad16, Bultink16} \\
    Single-qubit gate time                                  & $\Tgone$                  & $20~\ns$              &~\cite{Asaad16, Bultink16} \\
    Two-qubit gate time                                     & $\Tgtwo$                  & $40~\ns$              &~\cite{Riste15}            \\
    Coherent step time                                      & $\Tcorr$                  & $200~\ns$             &~\cite{Versluis16}         \\
    Measurement time                                        & $\Tmeas$                  & $300~\ns$             &~\cite{Bultink16}          \\
    Depletion time                                          & $\Tdep$                   & $300~\ns$             &~\cite{Bultink16}          \\
    Fast measurement time                                   & $\Tmeas^{\mathrm{(fast)}}$& $100~\ns$             &~\cite{Walter17}           \\
    Fast depletion time                                     & $\Tdep^{\mathrm{(fast)}}$  & $100~\ns$             &~\cite{Walter17} \\
    \hline
\end{tabular}
\caption{\label{table:parameters} Standard simulation parameters: Summary of standard times used in all density-matrix simulations, unless otherwise indicated. The two-qubit gate is a conditional phase gate (C-Z). Other error rates and parameters are given in~\cite{Suppmaterial}.}
\end{table}

\subsubsection{Idling qubits}
\label{sec:rest}
While idling for a time $\tau$, a transmon in $\ket{1}$ can relax to $\ket{0}$. Furthermore, a transmon in superposition can acquire random quantum phase shifts between $\ket{0}$ and $\ket{1}$ due to $1/f$ noise sources (e.g., flux noise) and broadband ones (e.g., photon shot noise~\cite{Sears12} and quasiparticle tunneling~\cite{Riste13}).
These combined effects can be parametrized by probabilities $p_{1}=\exp(-\tau/\Tone)$ for relaxation, and $p_{\phi}=\exp(-\tau/\Tphi)$ for pure dephasing.
The combined effects of relaxation and pure dephasing lead to decay of the off-diagonal elements of the qubit density matrix.
We model dephasing from broadband sources in this way, taking for $\Tphi$ the value extracted from the decay time $\Ttwo$ of standard echo experiments:
\begin{equation}
\frac{1}{\Ttwo}=\frac{1}{\Tphi}+\frac{1}{2\Tone}.
\end{equation}
We model $1/f$ sources differently, as discussed below.

\subsubsection{Dephasing from photon noise}
\label{sec:photons}
The dominant broadband dephasing source is the shot noise due to photons in the readout resonator.
This dephasing is present whenever the coupled qubit is brought into superposition before the readout resonator has returned to the vacuum state following the last measurement.
This leads to an additional, time-dependent pure dephasing (rates given in~\cite{Suppmaterial}).

\subsubsection{One-qubit Y rotations}
\label{sec:sqgates}
We model $y$-axis rotations as instantaneous rotations sandwiched by idling periods of duration $\Tgone/2$. The errors in the instantaneous gates are modeled from process matrices measured by gate-set tomography~\cite{Blume13,Blume16} in a recent experiment~\cite{Rol16}.
In this experiment, the GST analysis of single-qubit gates also showed that the errors can mostly be attributed to Markovian noise.
For simplicity, we thus model these errors as Markovian.

\subsubsection{Dephasing of flux-pulsed qubits}
\label{sec:dephasing}
During the coherent step, transmons are repeatedly moved in frequency away from their sweetspot using flux pulses, either to implement a C-Z gate or to avoid one.
Away from the sweetspot, transmons become first-order sensitive to flux noise, which causes an additional random phase shift.
As this noise typically has a $1/f$ power spectrum, the largest contribution comes from low-frequency components that are essentially static for a single run, but fluctuating between different runs.
In our simulation, we approximate the effect of this noise through ensemble averaging, with quasi-static phase error added to a transmon whenever it is flux pulsed.
Gaussian phase errors with the variance (calculated in~\cite{Suppmaterial}) are drawn independently for each qubit and for each run.

\subsubsection{C-Z gate error}
\label{sec:two_qubit_gates}
The C-Z gate is achieved by flux pulsing a transmon into the $\ket{11}\leftrightarrow \ket{02}$ avoided crossing with another, where the $2$ denotes the second-excited state of the fluxed transmon.
Holding the transmons here for $\Tgtwo$ causes the probability amplitudes of $\ket{01}$ and $\ket{11}$ to acquire phases~\cite{DiCarlo09}.
Careful tuning allows the phase $\phi_{01}$ acquired by $\ket{01}$ (the single-qubit phase $\phi_{\oneq}$) to be an even multiple of $2\pi$, and the phase $\phi_{11}$ acquired by $\ket{11}$ to be $\pi$ extra.
This extra phase acquired by $\ket{11}$ is the two-qubit phase $\phi_{\twoq}$.
Single- and two-qubit phases are affected by flux noise because the qubit is first-order sensitive during the gate.
Previously, we discussed the single-qubit phase error.
In~\cite{Suppmaterial}, we calculate the corresponding two-qubit phase error $\delta \phi_{\twoq}$.
Our full (but simplistic) model of the C-Z gate consists of an instantaneous C-Z gate with single-qubit phase error $\delta \phi_{\oneq}$ and two-qubit phase error $\delta \phi_{\twoq}=\delta \phi_{\oneq}/2$, sandwiched by idling intervals of duration $\Tgtwo/2$.

\subsubsection{Measurement}
\label{sec:measurement}
We model qubit measurement with a black-box description using parameters obtained from experiment.
This description consists of the eight probabilities for transitions from an input state $\ket{i}\in\{\ket{0},\ket{1}\}$ into pairs ($m$,$\ket{o}$) of measurement outcome $m\in\{+1,-1\}$ and final state $\ket{o}\in\{\ket{0},\ket{1}\}$.
By final state we mean the qubit state following the photon-depletion period.
Input superposition states in the computational bases are first projected to $\ket{0}$ and $\ket{1}$ following the Born rule.
The probability tree (the butterfly) is then used to obtain an output pair $\left(m,\ket{o}\right)$.
These experimental parameters can be described by a six-parameter model (described in detail in~\cite{Suppmaterial}), consisting of periods of enhanced noise before and after a point at which the qubit is perfectly projected, and two probabilities $\ero^{\ket{i}}$ for wrongly declaring the result of this projective measurement.
In~\cite{Suppmaterial}, a scheme for measuring these butterfly parameters and mapping them to the six-parameter model is described. In experiment, we find that the readout errors $\ero^{\ket{i}}$ are almost independent of the qubit state $\ket{i}$, and so we describe them with a single readout error parameter $\ero$ in this work.

\begin{acknowledgments}
    We thank C.~C.~Bultink, M.~A.~Rol, B.~Criger, X.~Fu, S.~Poletto, R.~Versluis, P.~Baireuther, D.~DiVincenzo, B.~Terhal, and C.W.J. Beenakker for useful discussions. This research is supported by the Foundation for Fundamental Research on Matter (FOM), the Netherlands Organization for Scientific Research (NWO/OCW), an ERC Synergy Grant, and by the Office of the Director of National Intelligence (ODNI), Intelligence Advanced Research Projects Activity (IARPA), via the U.S. Army Research Office grant W911NF-16-1-0071. The views and conclusions contained herein are those of the authors and should not be interpreted as necessarily representing the official policies or endorsements, either expressed or implied, of the ODNI, IARPA, or the U.S. Government. The U.S. Government is authorized to reproduce and distribute reprints for Governmental purposes notwithstanding any copyright annotation thereon.
\end{acknowledgments}

\bibliographystyle{naturemag}
\bibliography{References_cQED}

\appendix

\section{Full circuit diagram for Surface-17 implementation}\label{app:full_circuit}
\begin{figure*}
\includegraphics[width=\textwidth]{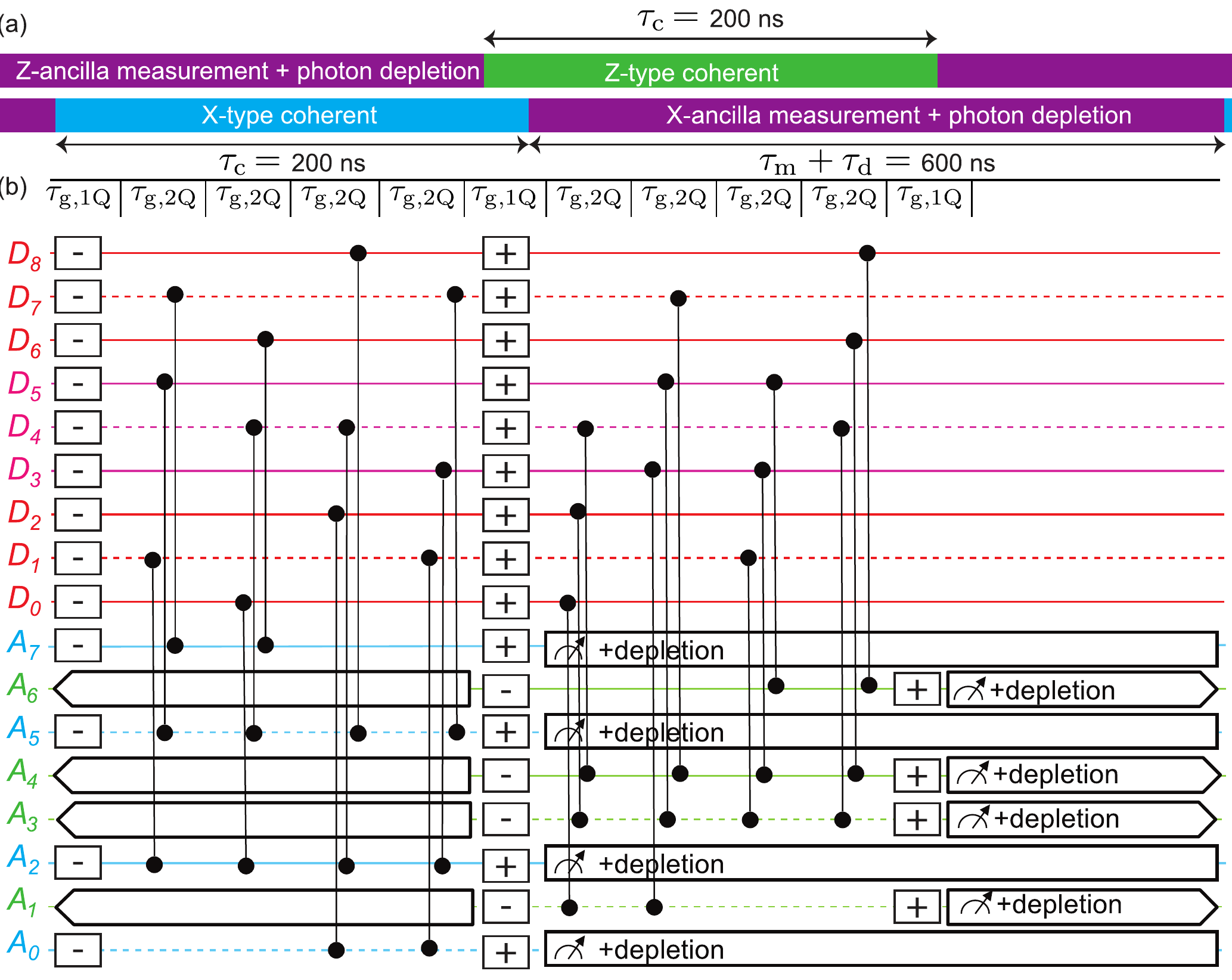}
\caption{\label{Fig_circuit}The quantum circuit for Surface-$17$ syndrome measurement used in all simulations.
(a) Outline of the timing of the standard circuit, including the time shift between X- and Z-type stabilizer measurements described by \cite{Versluis16}.
Qubit labels correspond to the position in Fig.~6.
(b) Full quantum circuit of the QEC cycle. The C-Z gates within each group are slightly offset horizontally for visibility (in reality they are performed simultaneously).
}
\end{figure*}

The quantum circuit \cite{Versluis16} (Fig.~\ref{Fig_circuit}) consists of $R_y(\pi/2)$ (``$+$'') and $R_y(-\pi/2)$ (``$-$'') rotations, C-Z gates, and ancilla measurements.
The coherent steps of the $X$ and $Z$ ancillas are pipelined (shifted in time with respect to each other) to prevent transmon-transmon avoided crossings. As long as $\Tmeas + \Tdep \geq \Tcorr$, no time is lost due to this separation.

\begin{figure}
\includegraphics[width=\columnwidth]{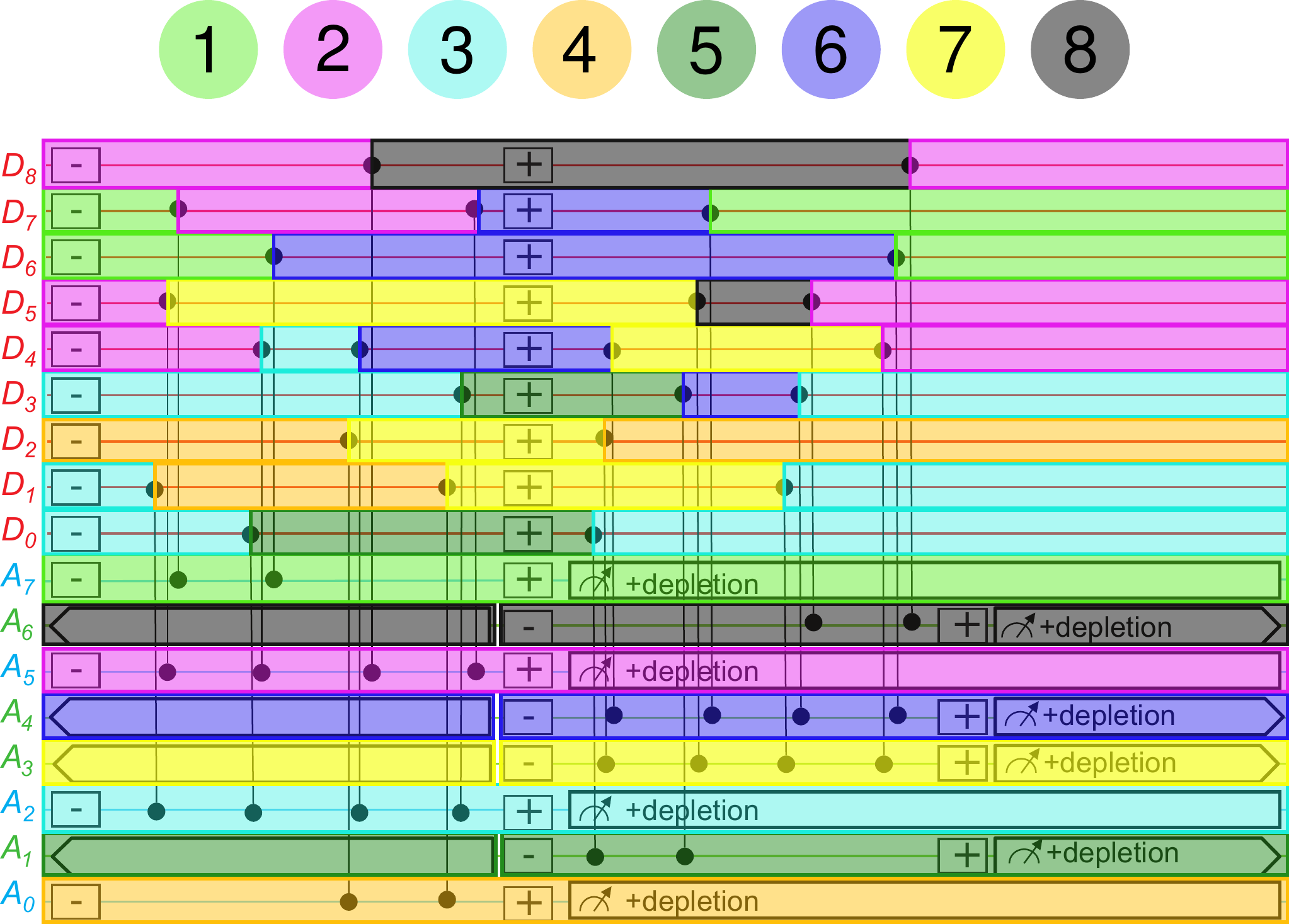}
\caption{\label{Fig_pipelined_circuit}Isolation of ancilla interactions in the Surface-17 circuit given in Fig.~\ref{Fig_circuit}. Throughout a simulation, quantumsim stores the density matrix of all data qubits. Each error correction cycle is split up into $8$ steps as labeled. In each step, a single ancilla qubit is added to the density matrix, the correspondingly colored pieces of the circuit are executed, and the ancilla is read out and removed from the density matrix. This scheme is only possible because on each data qubit all gates are executed in order. Note that steps after the final C-Z gate on a data qubit are executed during the next cycle.}
\end{figure}

In a simulation of the given circuit, gates on different qubits commute and may be applied to the density matrix in any order, regardless of the times at which they are performed in an experiment. As described in Sec.~IV~A~3, by simulating gates in a specific order (Fig.~\ref{Fig_pipelined_circuit}), 
one can ensure that only one ancilla is ancilla is entangled with the data qubits at any point in the simulation.
This allows a reduction in the maximum size of the density matrix from $2^{17}\times 2^{17}$ to $2^{10}\times 2^{10}$.

\section{Parameters of error models}\label{app:Pauli_transfer_matrices}
This appendix provides mathematical details of the sources of error described in the main text.
Standard values for the parameters used throughout the text are given in Table~\ref{table:parameters_extd}.

\begin{table}[h]
\begin{tabular}{| l | l | l|l| }
    \hline \hline
    Parameter               & Symbol                    & Value                 & Reference                 \\
    \hline
    In-axis rotation error      & $\paxis$                  & $10^{-4}$             &~\cite{Rol16}              \\
    In-plane rotation error     & $\pplane$                 & $5\times 10^{-4}$     &~\cite{Rol16}              \\
    $1/f$ flux noise.           & $A$                       & $(1\mu\Phi_0)^2$      &~\cite{Yan15,Quintana17}   \\
    Readout infidelity          & $\ero$                    & $5\times 10^{-3}$     &~\cite{Bultink16}          \\
    Photon relaxation time      & $1/\kphoton$              & $250~\ns$             &~\cite{Bultink16}          \\
    Dispersive shift            & $\disp$                   & $-2.6~\MHz$           &~\cite{Bultink16}          \\
    photon $\#$ post-measurement & $n_0$                & $0.8$ photons             &~\cite{Bultink16}          \\
    \hline \hline
\end{tabular}
\caption{\label{table:parameters_extd} Standard parameters of error models used in quantumsim, unless indicated otherwise.}
\end{table}

In the quantumsim module, all gates are applied in the Pauli transfer matrix representation~\cite{Chow12}. These are given in the form
\begin{equation}
{\left(R_{\Lambda}\right)}_{ij}=\frac{1}{2}\text{Tr}\left(\sigma_i\Lambda \sigma_j\right),
\end{equation}
where matrices $\sigma_i$ are the Pauli operators: $\sigma_0=I$, $\sigma_1=X$, $\sigma_2=Y$ and $\sigma_3=Z$.

\subsection{Qubit idling}
Idling qubits are described by the amplitude-phase damping model \cite{Nielsen00}, corresponding to the transfer matrices
\begin{eqnarray}
R_{\Lambda_{T_1}}&=&\left(\begin{array}{cccc}1&0&0&0\\0&\sqrt{1-p_{1}}&0&0\\0&0&\sqrt{1-p_{1}}&0\\p_{1}&0&0&1-p_{1}\end{array}\right)\label{T_1_matrix}\\
R_{\Lambda_{T_{\phi}}}&=&\left(\begin{array}{cccc}1&0&0&0\\0&\sqrt{1-p_{\phi}}&0&0\\0&0&\sqrt{1-p_{\phi}}&0\\0&0&0&1\end{array}\right)\label{T_phi_matrix}.
\end{eqnarray}
Idling for a duration $t$ is thus described by
\begin{equation}
R_{AP(t)} = R_{\Lambda_{T_1}}R_{\Lambda_{T_{\phi}}}
\end{equation}
with $p_1 = 1 - e^{-t/\Tone}$ and $p_\phi = 1 - e^{-t/\Tphi}$.
\subsection{Photon decay}\label{app:photons}
In the presence of photons in a readout resonator, the coupled qubit is affected according to the effective stochastic master equation~\cite{FriskKockum12}:
\[
\frac{d\rho}{dt}=-i\frac{B}{2}[\sigma_z,\rho]+\frac{\Gamma_d}{2}\mathcal{D}[\sigma_z]\rho.
\]
Here, $\rho$ is the qubit density matrix, $\mathcal{D}[X]$ is the Lindblad operator $\mathcal{D}[X]\rho=X\rho X^{\dag}-\frac{1}{2}X^\dag X\rho-\frac{1}{2}\rho X^\dag X$, $B=2\chi\text{Re}(\alpha_g\alpha_e^*)$ is the measurement-induced detuning (Stark shift), and $\Gamma_d=2\chi\text{Im}(\alpha_g\alpha_e^*)$ is the measurement-induced dephasing, with $\alpha_i$ the qubit-state-dependent photon field in the resonator and $2\chi$ the qubit frequency shift per photon.
At time $t-t_g$ after the qubit superposition is created,
\[
\alpha_g\alpha_e^*=\alpha(t_m)\exp\left(-\kappa\left(t-t_m\right)\right)\exp\left(2i\chi\left(t-t_g\right)\right),
\]
with $t-t_m$ the time since the end of measurement excitation pulse. Integrating over the interval $\left[t_1,t_2\right]$ gives
a dephasing term with coefficient
\[
\begin{array}{lll}
\pphoton&=&\exp\left(-\int_{t_1}^{t_2}\Gamma_d(t) dt\right)\\
    &=& \exp\left(\vphantom{{\left[\frac{e^{-\kappa t}}{4\chi^2+\kappa^2}[-\kappa\sin(2\chi t)-2\chi\cos(2\chi t)]\right]}^{t_2-t_g}_{t_1-t_g}}2\chi\alpha(0)\exp(\kappa(t_m-t_g))\right. \\
    & & \times\left.{\left[\frac{e^{-\kappa t}}{4\chi^2+\kappa^2}[-\kappa\sin(2\chi t)-2\chi\cos(2\chi t)]\right]}^{t_2-t_g}_{t_1-t_g}\right).
\end{array}
\]
This dephasing is then implemented via the same Pauli transfer matrix as~\eqref{T_phi_matrix}.

\subsection{Single-qubit $R_y(\pi/2)$ rotations}\label{app:sq_rotations}

Single-qubit rotations are modeled by sandwiching an instantaneous Pauli transfer matrix, representing
the rotation, with periods of duration $\Tgone/2$ of amplitude and phase damping. This allows to model the gate for different $\Tone$ and $\Tphi$. However, comparison of this
model with Pauli transfer matrices obtained from gate-set tomography experiments
shows that actual gates are more accurately described when adding a phenomenological
depolarizing noise to the instantaneous part. In the Bloch sphere, this decay
corresponds to shrinking toward the origin, with factor $1-\paxis$ along the $y$ axis and
$1-\pplane$ along the $x$- and $z$-axes. We thus model
\begin{equation}
R_{R_y(\pi/2)} = R_{AP(\Tgone/2)} R^\prime_{R_y(\pi/2)} R_{\text{dep}} R_{AP(\Tgone/2)}\label{eq:rotation},
\end{equation}
where
$$
R_{\text{dep}} = \begin{pmatrix}
  1 & 0 & 0 & 0 \\
  0 & 1-\pplane & 0 & 0 \\
  0 & 0 & 1-\paxis & 0 \\
  0 & 0& 0 &1-\pplane \\
  \end{pmatrix},
$$
and $R^\prime_{R_y(\pi/2)}$ is the Pauli transfer matrix describing a perfect $\pi/2$ rotation around the $y$ axis.

\subsection{Flux noise}\label{app:flux_noise}
Shifting the transmon from its sweetspot $\fqmax$ to a lower frequency
\[
\fq(t)=(\fqmax+\EC) \sqrt{ \lvert \cos\left(\pi\Phi(t)/\Phio\right)\rvert}-\EC
\]
makes it first-order sensitive to flux noise, with sensitivity
\[
\frac{\partial \fq}{\partial \Phi}=\frac{-\pi}{2\Phio}(\fq+\EC)\tan \left( \frac{\pi \Phi}{\Phio}\right).
\]
Here, $\Phi$ is the flux bias and $\Phio=h/2e$ is the flux quantum.
For a deviation of $\delta \Phi$, the pulsed transmon incurs a phase error
\[
\delta \phi=- 2\pi\Tgtwo \frac{\partial \fq}{\partial \Phi} \delta \Phi.
\]
Flux noise has a characteristic (single-sided) spectral density
\[
S_{\Phi}(f) \approx A/f,
\]
where $A\approx {(1~\mu \Phio)}^2$ with $f$ in $\Hz$.
We model this noise as quasi-static over the duration ($1/f_{\min}\sim 20~\us$, or 20 QEC cycles) of individual runs, but fluctuating between subsequent runs ($1/f_{\max}\sim 20~\sec$, or $10^5$ runs at $200~\us$ intervals).
The root-mean-square (rms) fluctuations of flux are therefore
\[
\begin{array}{lcl}
    \delta \Phi_{\rms}  &=&{\left(\int_{f_{\min}}^{f_{\max}} S_\Phi(f)\,df \right)}^{1/2}\\
                    &=& A{\left( \ln \left( f_{\max} / f_{\min}\right)\right)}^{1/2}\\
                    &\approx & 4~\mu\Phio.
\end{array}
\]
For our quantum circuit based on \cite{Versluis16}, we estimate the corresponding rms phase error induced in a pulsed transmon to be
\[
\delta \phi_{\rms} \approx 0.01~\rad.
\]

\subsection{C-Z gates}\label{app:2q_flux_noise}
We now focus on the two-qubit phase error.
For an adiabatic gate,
\[
\phi_{\twoq} = \phi_{11}-\phi_{01}=-2 \pi \int_{t_1}^{t_2}\zeta(t)dt,
\]
with $t_1$ and $t_2=t_1+\Tgtwo$ the start and end of the gate and $\zeta$ the time-dependent frequency deviation of the lower branch of the $\ket{11}\leftrightarrow \ket{02}$ avoided crossing from the sum of frequencies for $\ket{01}$ and $\ket{10}$.
Near the flux center $\Phic$ of the $\ket{11}-\ket{02}$ avoided crossing,
\[
\zeta \approx \beta(\Phi-\Phic)-\sqrt{\beta^2{\left(\Phi-\Phic\right)}^2+{\left(2J/2\pi\right)}^2},
\]
where $2J/2\pi\sim50~\MHz$ is the minimum splitting between $\ket{11}$ and $\ket{02}$, and
\[
\beta=\frac{1}{2}\frac{\partial\fq}{\partial\Phi}|_{\Phi=\Phic}.
\]
Differentiating with respect to $\Phi$ at $\Phic$ gives
\[
\frac{\partial\zeta}{\partial\Phi}|_{\Phi=\Phic}=\beta.
\]
To estimate the $\delta \phi_{\twoq}$ error, we make the following simplification: we replace the exact trajectory created by the flux pulse by a shift to $\Phi=\Phic+\delta \Phi$ with duration $\Tgtwo$. For a deviation of $\delta \Phi$,
\[
\delta \phi_{\twoq}  \approx - 2\pi \Tgtwo  \frac{\partial\zeta}{\partial\Phi}|_{\Phi=\Phic} \delta \Phi.
\]
Note that this two-qubit phase error is correlated with the single-qubit phase error on the fluxed transmon. The former is smaller by a factor $\approx2$.

\subsection{Measurement}\label{app:measurement}

The probabilities $\butterfly{i}{m}{o}$ are calibrated using the statistics of outcomes in back-to-back measurements ($a$ followed by $b$) with the qubit initialized in $\ket{i}$.
\[
\begin{array}{lcl}
\mathrm{P}{(\meas{a}=+1)}_i &=& \butterfly{i}{+1}{0}+\butterfly{i}{+1}{1},\\
\mathrm{P}{(\meas{a}=+1)}_i &=& \butterfly{i}{-1}{0}+\butterfly{i}{-1}{1},\\
\mathrm{P}{(\meas{b}=\meas{a}=+1)}_i      &=&    \left(\butterfly{0}{+1}{0}+\butterfly{0}{+1}{1}\right)\butterfly{i}{+1}{0}\\
&&+ \left(\butterfly{1}{+1}{0}+\butterfly{1}{+1}{1}\right) \butterfly{i}{+1}{1}, \\
\mathrm{P}{(\meas{b}=-\meas{a}=+1)}_i     &=&   \left(\butterfly{0}{+1}{0}+\butterfly{0}{+1}{1}\right)\butterfly{i}{-1}{0} \\
&&+ \left(\butterfly{1}{+1}{0}+\butterfly{1}{+1}{1}\right) \butterfly{i}{-1}{1},\\
\mathrm{P}{(-\meas{b}=\meas{a}=+1)}_i     &=&   \left(\butterfly{0}{-1}{0}+\butterfly{0}{-1}{1}\right)\butterfly{i}{+1}{0} \\
&&+ \left(\butterfly{1}{-1}{0}+\butterfly{1}{-1}{1}\right) \butterfly{i}{+1}{1},\\
\mathrm{P}{(-\meas{b}=-\meas{a}=+1)}_i    &=&  \left(\butterfly{0}{-1}{0}+\butterfly{0}{-1}{1}\right)\butterfly{i}{-1}{0} \\
&&+ \left(\butterfly{1}{-1}{0}+\butterfly{1}{-1}{1}\right) \butterfly{i}{-1}{1}.
\end{array}
\]
We obtain the six free parameters of the black-box description from these 12 equations, using experimental values on the left-hand side~\cite{Riste12}.
Table~\ref{table_butterfly} shows the values used, achieved in a recent experiment~\cite{Bultink16}.
\begin{table}
\begin{tabular}{| l | l | l|l| }
    \hline \hline
    Probability             & Value  & Probability             & Value     \\
    \hline
    $\butterfly{0}{+1}{0}$  & 0.9985  &  $\butterfly{1}{+1}{0}$ &   0.0050      \\
    $\butterfly{0}{+1}{1}$  & 0.0000  &  $\butterfly{1}{+1}{1}$ &   0.0015     \\
    $\butterfly{0}{-1}{0}$  & 0.0015  &  $\butterfly{1}{-1}{0}$ &   0.0149      \\
    $\butterfly{0}{-1}{1}$  & 0.000  &  $\butterfly{1}{-1}{1}$ &    0.9786       \\
    \hline \hline
\end{tabular}
\caption{Measurement butterfly matching a recent characteristic experiment~\cite{Bultink16} using a Josephson parametric amplifier~\cite{Castellanos-Beltran08} in phase-preserving mode as the front end of the readout amplification chain.}
\label{table_butterfly}
\end{table}
\begin{figure}
  \includegraphics[width=\columnwidth]{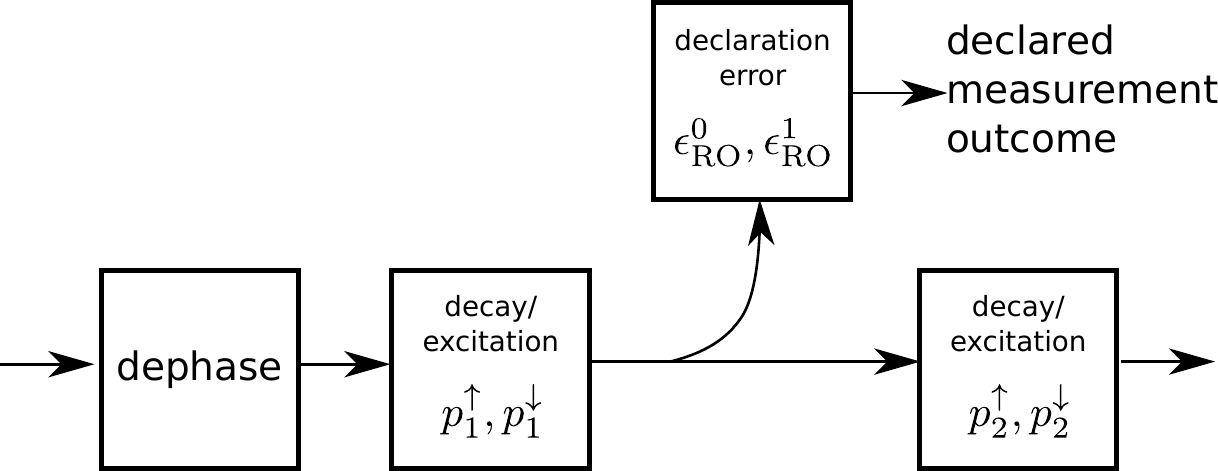}
  \caption{\label{fig:measurement_model}The model for measurements consists
      of a dephasing of the qubit followed by a period of decay and excitation
      with probability $p^{(1)}_{\downarrow/\uparrow}$. At this point, the
  qubit state is sampled. The sampling result is subject to a declaration error
  $\epsilon_{\mathrm{RO}}$, and the qubit state is subject to further decay or
  excitation with probabilities $ p^{(2)}_{\downarrow/\uparrow}$ before the end of the measurement block.}
\end{figure}
  For the simulation, we reproduce this behaviour of the measurement process by a model with several steps. 
  The qubit undergoes dephasing, followed by
periods of decay or excitation between which the measurement result
is sampled. This measurement result is further subject to a state-dependent
declaration error $\epsilon_{\mathrm{RO}}$ before reported to the decoder (see Fig.\ref{fig:measurement_model}).
The six parameters of this model are in a one-to-one correspondence with the butterfly parameters described above,
and can be mapped by solving the corresponding system of equations.
The experimental results in  Tab.\ref{table_butterfly} are very well
explained by assuming
unmodified amplitude-phase damping (withe zero excitation probabilities) during
the measurement period, and an outcome-independent declaration error of 
$\ero = \ero^1 =  \ero^0 =  0.15\%$. We use this result to extrapolate measurement
performance to different values of $\Tone$.

Reduction of measurement time is expected to reduce assignment fidelity. For
the results presented in Fig.~2, we do not rely on experimental results, but assume a simplified model
for measurement, following Ref.~\onlinecite{FriskKockum12}. A constant drive
pulse of amplitude $\epsilon$ and tuned to the bare resonator frequency,
$\Delta_r = 0$, excites the readout resonator for time $\Tmeas$. The dynamics
of the resonator is dependent on the transmon state (we approximate linear
behavior), and the transmitted signal is amplified and detected in a homodyne
measurement as a noisy transient.  This transient is processed by a linear
classifier, which declares the measurement outcome.  For resonator depletion,
we use a two-step clearing pulse with amplitude $\epsilon_{c1}$ and
$\epsilon_{c2}$, each active for $\Tdep/2$ and chosen (by numerical
minimization) so that, at the end of the depletion pulse, the transients for
both transmon states return to zero.
While the resonator dynamics is easily found if the transmon is in the ground
state, amplitude damping of the transmon in the excited state leads to
non-deterministic behavior. We thus numerically obtain an
ensemble of noisy transients for each input qubit state, and optimize the decision boundary of the linear classifier for this ensemble.
Generating a second verification ensemble, the ``butterfly'' of the measurement setup is estimated.

The dynamics of the resonator is determined by the resonator linewidth $\kappa$ as well as the dispersive shift $\chi$. We chose the parameters of the setup used in \cite{Bultink16}, $1/\kappa = 250\,\text{ns}$ and $\chi/\pi = -2.6\,\text{MHz}$. The signal-to-noise ratio of the detected transient is reduced by the quantum efficiency $\eta = 12.5 \%$. The driving strength $\epsilon$ is chosen to
approximate the ``butterfly'' used in most of the main text, and corresponds to a steady-state average photon population of about $\bar{n} = 15$.
We then keep $\epsilon$ constant while changing the measurement time, keeping $\Tmeas = \Tdep$, to obtain the butterflies used in the density matrix simulation.
We ignore effects leading to measurement-induced mixing and non-linearity of the readout resonator.
Finally, since these simulations do not allow to make a realistic prediction about residual photon numbers achievable in experiments, we ignore this effect when using these results.

\section{Effect of over-rotations and two-qubit phase noise on logical error rate}
\begin{figure}
\includegraphics[width=\columnwidth]{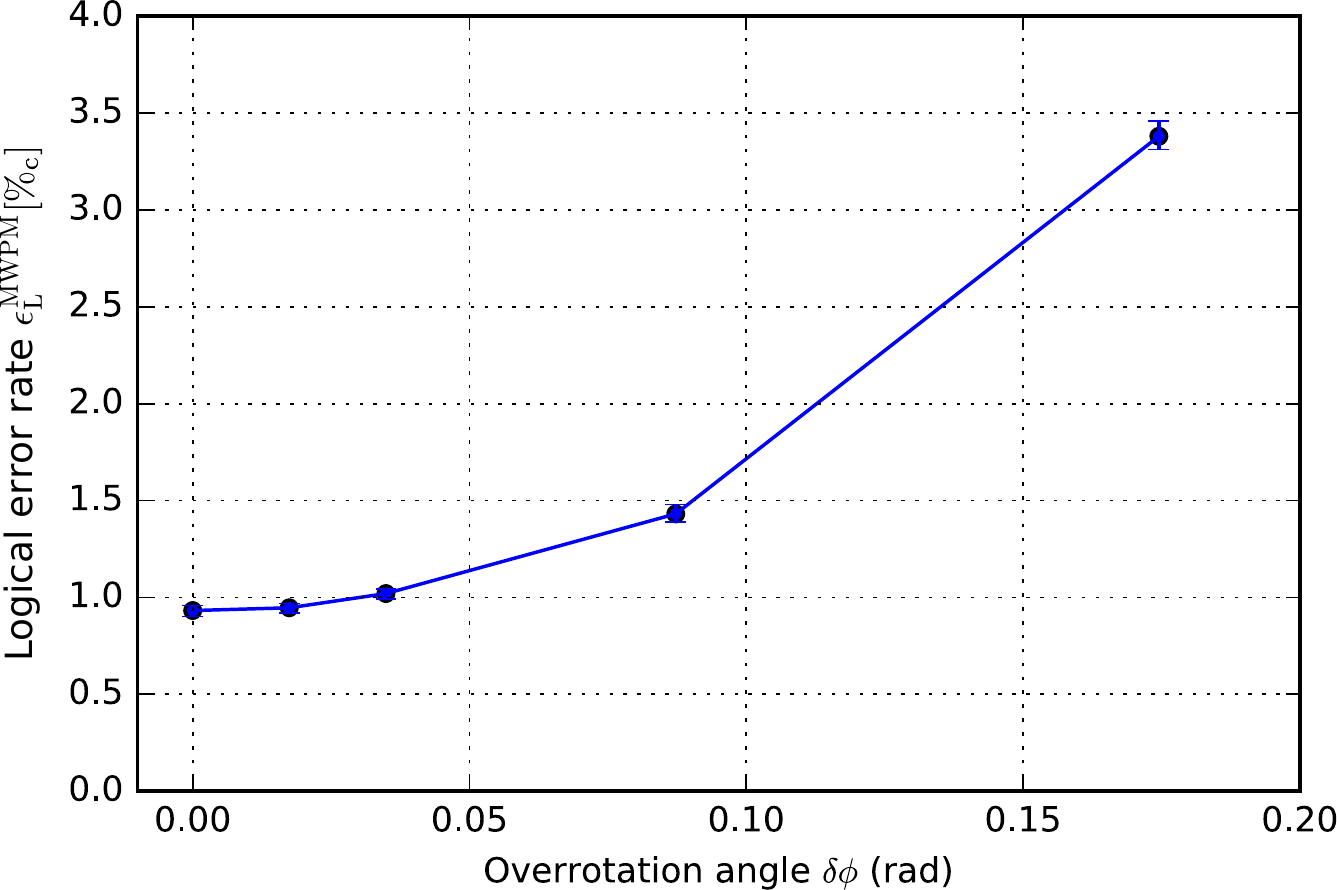}
\caption{\label{fig:err_overrotation}Logical error rate for Surface-$17$ as a function of single-qubit over-rotation, using the MWPM decoder. Other parameters are as given in the main text.}
\end{figure}

\begin{figure}
\includegraphics[width=\columnwidth]{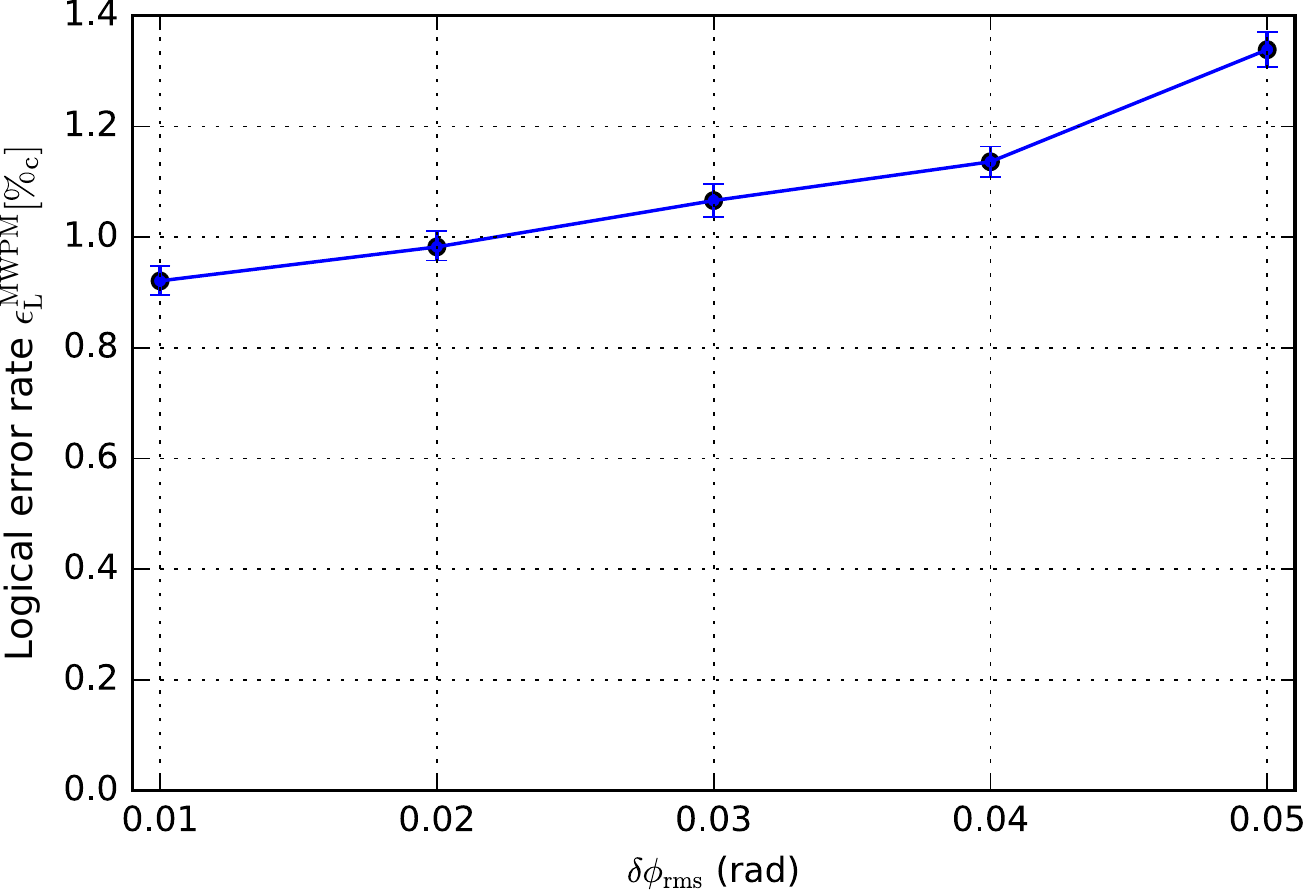}
\caption{\label{fig:err_stddev}Logical error rate for Surface-$17$ as a function of two-qubit phase error, using the MWPM decoder. Other parameters are as given in the main text.}
\end{figure}

In this section we provide additional numerical data showing the effect of some common noise sources on the logical error rate.
In Fig.~\ref{fig:err_overrotation} we show the effect of a coherent over-rotation, whereby the $R'_Y(\pi/2)$ operator in Eq.~\ref{eq:rotation} is replaced by $R'_Y(\pi/2+\delta\phi)$.
This can be caused by inaccurate calibration of the flux pulse used to perform the gate.
In Fig.~\ref{fig:err_stddev} we show the effect of an increase in the two-qubit flux noise $\delta\phi_{\mathrm{rms}}$ as described in Sec.~\ref{app:flux_noise}.

\section{Calculation of decoder upper bound}
We provide a detailed description how the decoder upper bound is obtained from the simulation results.
As described in the main text, after each cycle of simulation, the diagonal of the reduced density matrix of the data qubits in the Z basis is stored.
It contains the probability distribution for the $2^9 = 512$ different possible measurement outcomes of the data qubits.
In the quantum memory experiment described in the main text, each of these outcomes are passed to the decoder, which then declares a logical measurement outcome.

It is evident that any decoder must declare opposite logical outcomes if two of the 512 possible measurements m and m' are related by the application of a logical X operator.
Thus, any decoder can give the correct result only for half of the measurement outcomes.
Subject to this constraint, we can find the set of $256$ declarations which maximize the probability that the declaration is correct.
It immediately follows that no decoder can achieve a declaration fidelity larger than this maximal probability.
We thus refer to it as the decoder upper bound.

In practice, the upper bound is found according to the following approach.
Since declarations are opposite if two outcomes differ by a logical X operator, they must be equal if they differ by the application of one or more X stabilizers (applying two different logical X operators amounts to the application of a product of X stabilizers).
We thus group the outcomes in 32 cosets which are related by the application of X-stabilizers.
(There are 4 X-stabilizers in Surface-17, so there are $512/2^4 = 32$ cosets). 
For outcomes from the same coset, the declaration from a decoder must be the same. 
We obtain the probability of a final measurement falling within each coset by summing the probabilities from the density matrix diagonal.
We further group the 32 cosets to 16 pairs, which differ by the application of a logical operator.
The upper bound is then obtained by selecting the more probable coset from each pair and summing the corresponding probabilities.
This upper bound can also be interpreted as the internal decoherence of the logical qubit: it represents the maximal overlap of the final state with the initial state, under any possible correction of errors.

We finally emphasize that the this upper bound can be found only because we have access to the complete probability distribution of outcomes (for a given result of syndrome measurements), a major advantage of the density matrix simulation.
However, we do not expect that any decoder can actually achieve this upper bound: This is because we add syndrome measurement events independently after the situation, which will decrease the logical error rate further.

\section{Hardware requirements of simulation}
The simulations are performed using the quantumsim package~\cite{quantumsim_website}, which were developed by the authors for this work.
The package is accelerated by performing the density matrix manipulations on a GPU (graphics card).
The simulations for this work were performed on a NVidia Tesla K40 GPU, on which we observed runtimes of about 0.5 seconds for the simulation of a run of k=20 cycles (25 ms per QEC cycle).
We also had the opportunity to test the software on a more modern GPU (NVidia Tesla P100), observing about 15 ms per cycle, and on a consumer-grade GPU (NVidia Quadro M2000), observing about 40 ms per cycle.
By comparison, the CPU is mostly idle during the simulation, except for handling of input and output.
The memory requirements are modest for both CPU and GPU RAM.
They are dominated by the storage of the density matrices and amount to a few ten megabytes.

\section{Homemade MWPM decoder with asymmetric weight calculation}\label{app:weight_calculation}
Every QEC code requires a decoder to track the most likely errors consistent with a given set of stabilizer measurements.
The MWPM decoder has gained popularity since it was shown to have threshold values above $1\%$~\cite{Fowler09}.
The motivation behind MWPM is that single $X$ or $Z$ errors on data qubits in the bulk of a surface-code fabric cause changes of two stabilizers in the code. 
These signals can then be considered vertices on a graph, with the error the edge connecting them.
Errors in measurement, or errors on a single ancilla qubit, behave as changes in the stabilizer that are separated in time.
Multiple errors that would join the same vertices create longer paths in the graph, of which an experiment only records the endpoints.
Thus, the problem becomes that of finding the most likely set of generating errors given the error signals that mark their ends.
This is made slightly simpler, as in the surface code any chain of errors that forms a closed loop does not change the logical state.
This implies that all paths that connect two points are equivalent, and can be considered together.
The problem then is to join error signals, either in pairs, or to a `boundary' vertex. The latter corresponds to errors on data qubits at the boundary, which belong to only one $X$ or $Z$ stabilizer.
This pairing $P$ should be chosen as the most likely combination of single-qubit errors that could generate the measured error signals.
This has then been reduced to the problem of minimum-weight perfect matching on a graph, which can be solved in polynomial time by the blossom algorithm~\cite{Edmonds65,Fowler12}.

The MWPM decoder we use differs from previous methods by its weight calculation.
As part of the decoding process, it is required to calculate to some degree of accuracy~\cite{Fowler12b} the probability $p_{e_1,e_2}$ of two measured error signals $e_1$ and $e_2$ being connected by a chain of individual logical errors.
This is then converted to a weight $w_{e_1,e_2}=-\log(p_{e_1,e_2})$, which form the input to the blossom algorithm of Edmonds to find the most likely matching of error signals~\cite{Edmonds65,Fowler12}.
An exact calculation of $p_{e_1,e_2}$ requires a sum over all such chains between $e_1$ and $e_2$ that do not cross the boundary (these are equivalent modulo stabilizer operators that do not change the logical state).
In this appendix we detail a method of computing this sum, and approximations to make it viable within the runtime of the experiment.

Let us define the ancilla graph $\GA=(\VA,\EA)$ containing a vertex $v\in \VA$ for every ancilla measurement, and an edge $e\in \EA$ connecting $v,u\in V_A$ if a single component (gate, single-qubit rest period, or faulty measurement) in the simulation can cause the $u$ and $v$ measurements to return an error.
We include a special `boundary' vertex $v_B$, to which we connect another vertex $v$ if single components can cause errors on $v$ alone.
Then, to each edge $e$ we associate a probability $p_e$, being the sum of the probabilities of each component causing this error signal.
These error rates can be obtained directly from quantumsim, by cutting the circuit at each C-Z gate and measuring the decay of single qubits between.
Then, for a given experiment with given syndrome measurements, let us define the syndrome graph $\GS=(\VS,\ES)$ containing a vertex $v\in \VS$ for each syndrome measurement that records an error, and an edge $\lambda_{u,v}\in \ES$ connecting $u,v\in \VS$ if $u$ and $v$ are either both $X$ ancilla qubits or both $Z$ ancilla qubits.
To each edge $\lambda_{u,v}$ we associate a probability $p_{u,v}$ given by the sum of the probabilities of a chain of errors causing error signals solely on $u$ and $v$.

If we assume that single-qubit errors are uncorrelated, we have to lowest order
\begin{equation}
    p_{u,v}\approx\sum_{\text{paths } (e_1,e_2,\ldots,e_n)\\ \text{ between }u\text{ and }v}\quad\prod_{j=1}^n p_{e_j},
\end{equation}
Let $\AA$ be the adjacency matrix on $\GA$ weighted by the probabilities $p_e$ (i.e., $(\AA)_{u,v}=p_e$ with $e$ connecting $u$ and $v$), and $\AS$ the same for $\GS$. Then, the above becomes
\begin{equation}
\AS=\AA+\AA^2+\AA^3+\cdots=\frac{\mathbf{1}}{\mathbf{1}-\AA}-\mathbf{1},\label{eq:invert_eqn}
\end{equation}
noting that $\AS$ contains a subset of the indices that are used to construct $\AA$.

The boundary must be treated specially in the above calculation.
For the purposes of the surface code, the boundary can be described as a single vertex which has no limit on the number of other vertices it may pair to~\cite{Fowler12}.
For the purposes of weight calculation, any path that passes through the boundary is already counted by pairing both end vertices to the boundary.
This can be treated by making $\GA$ directed, and breaking the symmetry $\AA^T=\AA$.
In particular, either ${(\AA)}_{v_B,u}=0$ for all $u$ or ${(\AA)}_{u,v_B}=0$ for all $u$.

The above calculation requires inversion of a $\Nmat\times \Nmat$ matrix, with $\Nmat$ the total number of ancilla measurements per experiment.
Furthermore, as ancilla error rates depend upon the previous ancilla state, elements in $\AA$ are not completely known until the previous cycle. This implies that in an actual computation with runtime decoding, this inversion would need to be completed within a few microseconds (with a transmon-cQED architecture), which is practically unfeasible.
We suggest two approximations that can be made to shorten the decoding time.
The first is to average all errors over the ancilla population, ignoring any asymmetry in the system.
The adjacency matrix is now the same for any experiment, and can be precalculated and stored as a look-up table for the run-time decoder.
We call this the decoder with symmetrized weights.
The size of such a look-up table scales poorly with the number of qubits and the number of cycles.
However, $(\AS)_{u,v}$ is approximately invariant under simultaneous translation of $u$ and $v$ (excluding boundary effects).
This implies that a precalculated $\AS$ can be vastly compressed, making this method feasible.

The second approximation to the full $\AS$ calculation is to perform it iteratively.
We divide our graph $\GA$ ($\GS$) by time steps; let $\GA^t$ ($\GS^t$) be the subgraph of $\GA$ ($\GS$) containing only ancillas measured before time step $t$, and let $\partial\GA^t$ ($\partial\GS^t$) be the subgraph of $\GA$ ($\GS$) containing only ancillas measured during time step $t$.
Then, if we assume we have an approximation to the matrix $\AS^t$ (being the adjacency matrix of $\GS^t$), we can approximate
\begin{equation}
\AS^{t+1}\approx\left(\begin{array}{cc}\AS^t&C_S^{t+1}\\{(C_{S}^{t+1})}^T&{(\mathbf{1}-\partial\AA^{t+1})}^{-1}\end{array}\right)
\label{eq:approx}
\end{equation}
to lowest order in physical errors. Here, $\partial\AA^{t+1}$ is the weighted adjacency matrix on $\partial\GA^{t+1}$, and the coupling matrix $C_S^{t+1}$ is approximated by
\begin{equation}
C_{S}^{t+1}=\AS^t C_A^{t+1}{(\mathbf{1}-\partial\AA^{t+1})}^{-1},
\label{eq:prod_approx}
\end{equation}
with $C_A^{t+1}$ the adjacency matrix containing only edges between $\partial\GA^{t+1}$ and $\GA^t$.
This procedure corresponds to a sum over all paths that are made by moving within $\partial\GA^{t+1}$, shifting back in time to $\GA^t$, and then taking any precalculated path in $\GA^t$.
$C_{A}^{t+1}$ and ${(\mathbf{1}-\partial\AA^{t+1})}^{-1}$ can be precalculated, and so the runtime computation requirement is reduced to the product in Eq.~\eqref{eq:prod_approx}.
This in turn can be sparsified, as $C_A^{t+1}$ only contains connections to vertices in $\GA^t$ close to the time boundary, and we can delete all terms in $\AS^t$ that do not connect from these vertices to errors.

\begin{figure}
\includegraphics[width=\columnwidth]{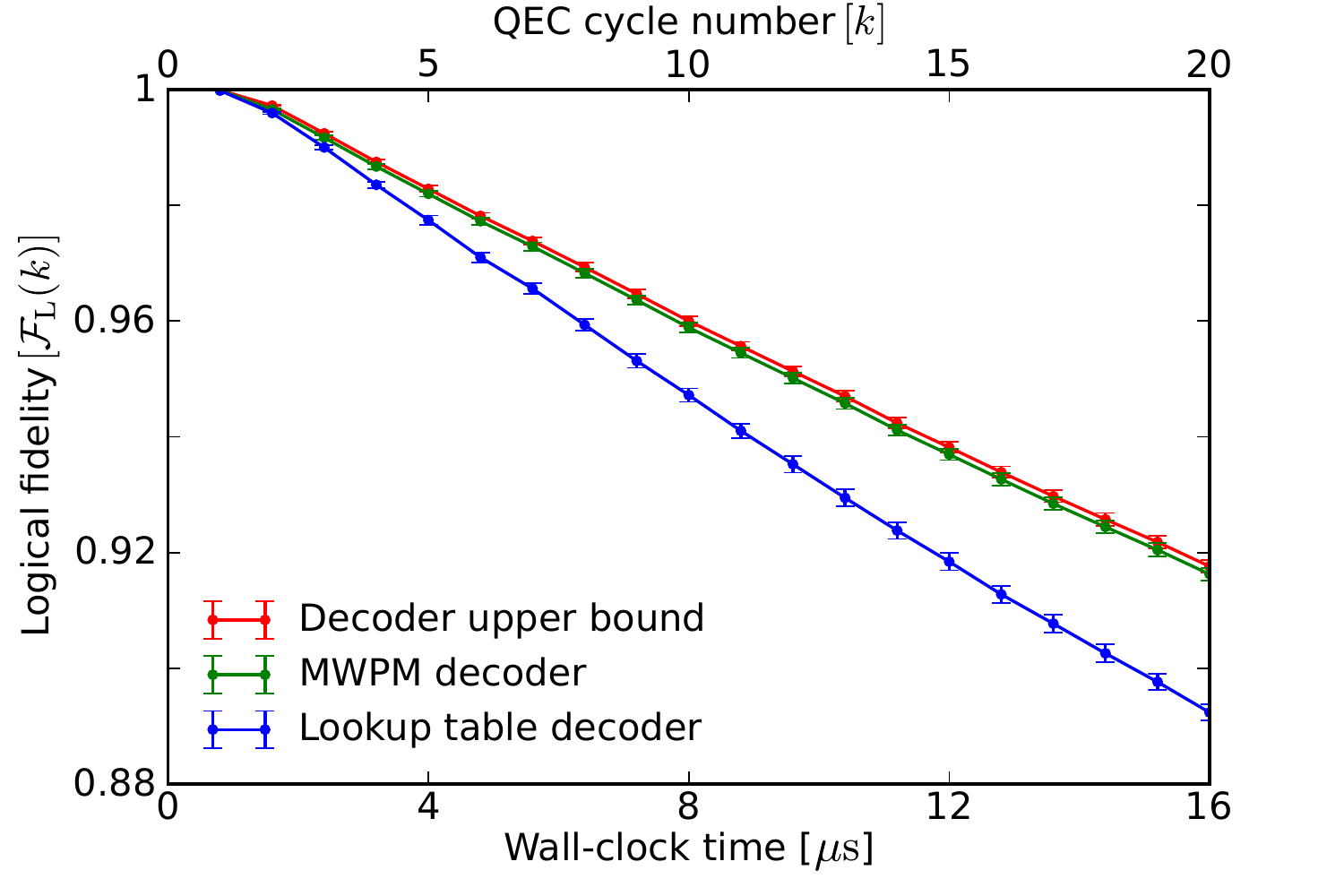}
\caption{\label{Fig_bitflip}Simulation of the experimental protocol used throughout the work, but using an error model that has $Y$ errors and readout infidelity removed. With these errors absent, the MWPM decoder achieves the decoder upper bound within simulation error. The look-up table approach (blue) retains some inaccuracy beyond this.}
\end{figure}

We have used the second method for our MWPM decoder, as we expect the error from neglecting higher-order combinations of errors to be small.
In order to check this assumption, in Fig.~\ref{Fig_bitflip} we repeat our simulation protocol with a modified physical error model that excludes all $Y$ and measurement errors.
We see that in the absence of these errors, the MWPM decoder performs within the error margin of the decoder upper bound.
Note that a small deviation is expected from the discrepancy between a MWPM decoder and a maximum-likelihood decoder~\cite{Heim16}.
With the parameters used in this work, we do not observe any loss of fidelity when we stop accounting for the difference in error rates between ancilla states.
We account this to the large error contribution from photon noise and gate infidelity on the ancilla qubits, which do not have this asymmetry.
We further note that we operate in a regime of large ancilla error; as described in the text this makes the system counter-intuitively less sensitive to ancilla noise.
In systems where this is not the case, it could be that accounting for ancilla asymmetry provides a useful computational method to improve $\eL$.

\section{Implementation of a look-up table decoder}\label{app:TomitaSvore}
In~\cite{Tomita14}, the authors describe a decoding scheme specific to Surface-$17$, which is optimized to be implementable with limited computational resources in a short cycle time.
This decoding scheme works by using a short decision tree to connect errors to each other in a style similar to blossom.
Indeed, this scheme is equivalent to a blossom decoder with all horizontal, vertical and diagonal weights equal~\cite{Tomita14}.
As such, we have implemented the new weights in the blossom decoder rather than utilizing the exact method given.

\section{Details of lowest-order approximation}\label{app:approx}
We detail the approximation made to study Surface-$49$ in Sec.~II~C. Note that this calculation is only for $X$ errors, which are measured by the $Z$ ancillas.
This implies that our approximation should attempt to realize the result of blossom, rather than the decoder upper bound.

We begin with the $\GA$ graph defined in App.~\ref{app:weight_calculation}.
In the absence of correlated errors that cause more than two error signals, any experiment can be approximately described by choosing a set $S\subset\EA$ of edges on the graph and assuming the errors that correspond to these edges have occurred.
Each ancilla measurement corresponds to a vertex in $\GA$, which records an error if an odd number of edges in $S$ point to the vertex.
Each combination $M_a$ of ancilla measurements can be generated by multiple error sets $S$.

Formally, let us write $\mathcal{M}$ for the set of all combinations of ancilla measurements and $\mathcal{S}$ for the set of all combinations of errors (so $\mathcal{S}=2^{\EA}$).
We then define a function $\phi:\mathcal{S}\rightarrow\mathcal{M}$ that takes a combination of errors to the resultant measurement outcomes.
Let us fix a logical $Z$ operator $Z_L$ on the surface-code fabric.
Then to each $S\in\mathcal{S}$ we can assign a parity $p(S)=\pm 1$ depending on whether the product of all errors in $S$ commute with $Z$ or not.
A decoding then consists of a choice of parity $p_d(M)$ for each $M\in\mathcal{M}$.
Such a decoding correctly decodes $S\in\mathcal{S}$ if $p_d(\phi(S))=p(S)$, and creates a logical error otherwise.
The source of logical errors in a perfect decoder is then precisely the fact that we can have two error combinations $S_1,S_2\in\mathcal{S}$ such that $\phi(S_1)=\phi(S_2)$ but $p(S_1)\neq p(S_2)$.

The above suggests a method by which a perfect decoder can be constructed.
As defined, $\phi^{-1}(M)\subset\mathcal{S}$ is the set of error combinations $S$ that return a measurement $M_\in\mathcal{M}$.
For each error combination $S$, we can calculate the probability of this occurring:
\begin{equation}
r(S)=\prod_{e\in S}p_e\prod_{e\notin S}(1-p_e).\label{eq:rdef}
\end{equation}
The optimal choice of $p_d(M_a)$ is the one maximizing
\begin{equation}
\sum_{S\in\phi^{-1}(M),p(S)=p_d(M)}r(S),
\end{equation}
and the fidelity of such a decoder (over the entire experiment) can be calculated as
\begin{align}
\fid=1-\sum_{M\in\mathcal{M}}\min\left(\sum_{S\in\phi^{-1}(M)}\delta_{p(S),+1}\; \;r(S)\right.,\nonumber\\
\left.\sum_{S\in\phi^{-1}(M)}\delta_{p(S),-1}\; \;r(S)\right)\label{prob_sum}.
\end{align}
At this point the only approximation that has been made is to neglect the $\Tone$ asymmetry in the system, which we have shown previously in this work to be negligible.
Unfortunately, the above function cannot be evaluated exactly; the number of error combinations $S$ is approximately $2^{200}$ for $4$ cycles of Surface-$49$.
Our goal instead is to approximate this to the lowest order in the physical qubit error rate.

Let us make the approximation that our error combinations $S$ can be split into small, well-separated pieces of errors containing separate correctable and non-correctable parts, $S=\cup_i S^i$.
To each $S^i$ we can assign a time step $t(S^i)$, being the earliest time of the first error measurement observed (in $\phi(S^i)$).
The error rate per round, $\eL$, can be determined by summing Eq.~\ref{prob_sum} over all pieces $S^i$ of all combinations $S$ such that $t(S^i)=T$ (with arbitrary $T$), as the effect of repeated errors from $S^i, S^j\subset S$ is taken into account during the derivation of the logical fidelity equation (Eq.~2 in the main text).
\begin{align}
\eL=\sum_{M\in\mathcal{M}}\min\left(\sum_{S\in\phi^{-1}(M)}\sum_{t(S^i)=T}\delta_{p(S^i),+1}\; \;r(S)\right.,\nonumber \\
\left.\sum_{S\in\phi^{-1}(M_)}\sum_{t(S^i)=T}\delta_{p(S^i),-1}\; \;r(S)\right).
\end{align}
Let us also extend the above division of $S$ to a division of $M$ into separate pieces $M^a$, and rewrite our sum slightly,
\begin{align}
\eL=\sum_{M^a}\;\min\left(\sum_{S^i\in\phi^{-1}(M^a),t(S^i)=T}\delta_{p(S^i),+1}\bar{r}(S^i)\right.,\nonumber \\
\left.\sum_{S^i\in\phi^{-1}(M^a),t(S^i)=T}\delta_{p(S^i),-1}\; \;\bar{r}(S^i)\right),\label{eq:approx_e}
\end{align}
Where here we have brought the sum over the global combinations of syndromes and measurements inside a new function $\bar{r}$
\begin{align}
\bar{r}(S^i)=\prod_{e\in S^i}p_e&\sum_{M\supset M^a}\sum_{(S\supset S^i,S\in\phi^{-1}(M))}\nonumber\\
\;&\prod_{f\in S/S^i}p_f\prod_{g\notin S^i}(1-p_g)\nonumber\\
=\prod_{e\in S^i}p_e&\sum_{S\supset S^i}\;\prod_{f\in S/S^i}p_f\prod_{g\notin S^i}(1-p_g)\label{eq:approx_r}
\end{align}
If we took this approximation literally and considered the sum over every possible combination $S$ containing $S^i$, the final sum in Eq.~\ref{eq:approx_r} would reduce to
\begin{equation}
\bar{r}^{(\mathrm{u})}(S^i)=\prod_{e\in S^i}p_e.\label{eq:adjusted_r2}
\end{equation}
However, this includes error combinations $S$ that cannot be easily separated into $S^i$ and `something else', i.e.~they contain other errors $e$ that cannot be separated from $S^i$.
Eq.~\ref{eq:adjusted_r2} is then equivalent to assuming that if $S^i$ is an uncorrectable logical error, no nearby combination of physical errors $S'$ can be combined such that $S^i\cup S'$ is correctable unless $S'$ itself is an uncorrectable logical error.
Such combinations would serve to reduce the calculated $\epsilon_L$, and so $\bar{r}^{(u)}$ gives an upper bound for $\epsilon_L$ in Eq.~\ref{eq:approx_e}.
For a lower bound, we approximate that for any uncorrectable error combination $S^i$, approximately one rounds-worth of single errors would undo the logical error, leading to the approximation
\begin{equation}
\bar{r}^{(\mathrm{l})}(S^i)=\prod_{e\in S^i}p_e\prod_{t(\{e\})=T}(1-p_e).\label{adjusted_r3}
\end{equation}
We now make one further approximation, and sum Eq.~\ref{eq:approx_e} only over the shortest $S^i$ that can be expected to contribute to the final error rate.
That is, we sum over those $S^i$ with $|S^i|\leq (d+1)/2$, and that spread directly across the chain.
The error incurred from this approximation is roughly proportional to the largest single error, which is no more than $5\%$ throughout our study.
We use $\bar{r}^{(u)}$ and $\bar{r}^{(l)}$ to give the error bars shown in Fig.~5.
Points in the plot are taken as a log average of the upper and lower bounds, and thus have no particular relevance themselves.
We see that the numerical calculation falls within the corresponding error bars for almost the entire dataset, giving verification for our method, save a slight deviation at one point where it falls below.
Moreover, as the simulated Surface-$17$ error rate lies above the upper bound found for the Surface-$49$ error rate (with the standard set of parameters from the main text), our claim that Surface-$17$ will operate below the fault-tolerant threshold is quite strong.

\end{document}